\documentclass{article}
\usepackage{misc/arxiv3}

\usepackage{amsmath}
\usepackage{url} 
\usepackage{amsthm} 
\usepackage[colorlinks=true,linkcolor=red,filecolor=green,citecolor=blue]{hyperref}
\usepackage{comment}
\usepackage{graphicx} 
\usepackage{caption}
\usepackage{subcaption}
\usepackage{algorithmic} 
\usepackage{url}
\PassOptionsToPackage{hyphens}{url}
\usepackage{hyperref}
\expandafter\def\expandafter\UrlBreaks\expandafter{\UrlBreaks\do\-\do\~\do\'\do\"\do\-}%

\usepackage{changepage}

\usepackage[dvipsnames]{xcolor} 

\definecolor{colour1}{RGB}{166,206,227}
\definecolor{colour2}{RGB}{31,120,180}
\definecolor{colour3}{RGB}{178,55,250} 
\definecolor{colour4}{RGB}{51,160,44}

\newcounter{noteMCctr} 

\newcommand{\mc}[1]{{\textcolor{black}{{{}}#1}}{}}

\newcommand{\mcr}[1]{{\textcolor{black}{{{}}#1}}{}}

\newcounter{noteGRctr} 

 \newcommand{\gr}[1]{{\textcolor{black}{{{}}#1}}{}}

 \newcommand{\gdr}[1]{{\textcolor{black}{{{}}#1}}{}}

\newcounter{noteAMctr} 

 \newcommand{\am}[1]{{\textcolor{black}{{{}}#1}}{}}

\RequirePackage[authoryear]{natbib}
\title{The GNAR-edge model: A network autoregressive model for networks with time-varying edge weights}

\author[1]{Anastasia Mantziou}
\author[1,2]{Mihai Cucuringu}
\author[3]{Victor Meirinhos}
\author[1,2]{Gesine Reinert}
\affil[1]{The Alan Turing Institute}
\affil[2]{University of Oxford, Department of Statistics}
\affil[3]{Office for National Statistics}

\begin{document}

\maketitle

\begin{abstract}
In economic and financial applications, there is often the need for analysing multivariate time series, comprising of time series for a range of quantities.
In some applications such complex systems can be associated with some underlying network describing pairwise relationships among the quantities. Accounting for the underlying network structure for the analysis of this type of multivariate time series is required for assessing estimation error and can be particularly informative for forecasting. 
Our work is motivated by a dataset consisting of time series of industry-to-industry 
transactions. In this example, pairwise relationships between Standard Industrial Classification (SIC) codes can be represented using a network, with SIC codes as nodes and pairwise transactions between SIC codes as edges, while the observed time series of the amounts of the transactions for each pair of SIC codes can be regarded as time-varying weights on the edges. Inspired by \cite{knight2019generalised}, we introduce the GNAR-edge model which allows modelling of multiple time series utilising the network structure, assuming that each edge weight depends not only on its past values, but also on past values of its neighbouring edges, for a range of neighbourhood stages. 
The method is validated through simulations. Results from the implementation of the GNAR-edge model on  the real industry-to-industry data show  good fitting and predictive performance of the model. The predictive performance is improved when sparsifying the network using  a lead-lag analysis and thresholding edges according to a lead-lag score.
\end{abstract}



\section*{Disclaimer}
The views expressed are those of the authors and may not reflect the views of the Office for National Statistics or the wider UK Government.

\section{Introduction}\label{sec1}

Multivariate correlated time series commonly arise in many contexts, such as in economics and finance. In some applications, such correlated time series can be represented by an underlying network with multivariate time series on the edges describing pairwise relationships among the quantities. This paper illustrates that network representations of this type can be informative for explaining the data through parameter estimation, as well as for forecasting. 

There are two common approaches for modelling multivariate time series systems.
The first, simple and fast approach is to model each time series individually, using an autoregressive model \cite{box2015time}. This approach lacks in leveraging information from the whole time series system jointly. The second approach is to jointly model a Vector Autoregressive (VAR) model \citep{lutkepohl2005new, tsay2013multivariate}. 
This approach uses information from the entire time series system; a key limitation is the complexity of the model with respect to the parameters under estimation, which increases quadratically with increasing number of time series (see Section \ref{sec3} for details).
In light of this limitation, a range of studies have focused on imposing sparsity on the coefficients of the classic VAR model, with the goal to allow statistical inference for multivariate time series in high-dimensional settings. Among the most commonly used shrinkage approaches are
Lasso-based regularization approaches, introducing a penalty in the optimisation problem for estimating the model parameters; the penalty provides
control over the number of parameters that are meaningful to be incorporated in each setting. Notably, in the recent study of \cite{nicholson2020high}, the authors introduce a Lasso-type approach for sparsifying the classic VAR model using a hierarchical lag group penalty function, called HLAG, that leverages the ordered structure of the lag coefficients in a VAR. Another recent study by \cite{dallakyan2022time} utilises time series graphical Lasso, extending the two-stage inference framework for sparse VAR models introduced by \cite{davis2016sparse}. Alternative methods for dimensionality reduction for the VAR model involve the use of information criteria \citep{tsay2013multivariate,lutkepohl2005new}, Bayesian methods \citep{koop2013forecasting} and factor models \citep{bernanke2005measuring}, to name a few. A concise review of the different sparsity approaches in statistics literature is presented in \cite{nicholson2020high}. With the Lasso-type approaches, the reduction in the number of parameters of the VAR model is performed in the inference step of the unrestricted VAR model. When the multivariate time series is known to have network structure, then this information can be utilised in advance in the VAR model, forming a restricted VAR model which is tailored to network time series, as we discuss next.

A commonly studied type of multivariate time series associated with a network structure, is a time series observed on the nodes of a network. 
Recent studies have focused on modelling such network time series, with the goal of a downstream prediction task.  \cite{zhu2017network} introduces the network vector autoregressive model (NAR), which is a special version of the classical VAR model that
incorporates network information explicitly. Specifically,  \cite{zhu2017network} assumes that the value of a node $i$ at time $t$ depends not only on previous values of the node, but also on previous values of nodes which are neighbours of node $i$, as well as a set of node-specific covariates. In this setting, \cite{zhu2017network}  considers only nodes which are
followed  by node $i$ via a directed edge.  \cite{knight2019generalised} extends this idea by introducing the generalized network autoregressive (GNAR) model that also incorporates the effect of larger neighborhood sizes for each node $i$. Both studies assume a fixed network structure across time. A network autoregressive model accounting for a {time-}varying network structure is proposed by \cite{kang2021dynamic}, who obtain a class of restricted, concatenated VAR models, and infer a dynamic network that changes for each VAR sub-model assumed. In a similar setup, \cite{spencer2015inferring} introduces an autoregressive model for nodal time series, restricted to the case of Directed Acyclic Graphs (DAG), while in the study of \cite{armillotta2021poisson} the authors propose a Poisson network autoregressive model for modelling count time series observed on the nodes of a network. Some special cases of network autoregressive models for nodal time series which extend the aforementioned studies, are proposed by \cite{zhu2019network} who 
develop a network quantile autoregressive model;  \cite{zhu2020multivariate} propose a different inferential framework for modelling social networks using a multivariate spatial autoregressive model,  and \cite{chen2022community}
incorporate  information about the community structure of the network in the autoregressive model. A slightly different approach to the aforementioned studies is that of \cite{sioofy2021network}; \gr{motivated by predicting COVID-19 infection, the authors}
develop a network autoregressive model for time series on nodes of star-shaped networks only.

Our work is related to the type of multivariate time series considered in the aforementioned studies, with the key difference that in our setting, the time series are observed on the edges of the network \am{as time-varying weights}, rather than the nodes. Specifically, we are motivated by a data set consisting of time series of industry-to-industry 
transactions, aggregated to Standard Industrial Classification (SIC) codes; SIC codes classify companies according to the nature of their business \cite{sic}. In this data, pairwise relationships between SIC codes can be represented using a network \gdr{with SIC codes as nodes}, while the observed time series for each pair of SIC codes can be regarded as time-varying edge weights. Inspired by the studies of \cite{zhu2017network} and \cite{knight2019generalised}, we introduce the GNAR-edge model that allows to capture network information by incorporating  the effect of neighbouring edges in the model. Similarly to the studies of \cite{zhu2017network} and \cite{knight2019generalised} and motivated by the real data application, we assume that the network structure is fixed over time.

There are a few related studies which take a network-based approach. One
class of class of studies \gr{which is} related to our work is the analysis of dynamic networks, \gdr{which} focuses on modelling the evolving structure of a network. In the dynamic networks literature, various models have been developed that extend models from the static network literature, such as Exponential Random Graphs (ERGM) \cite{krivitsky2014separable,hanneke2010discrete}, latent space models \cite{durante2016locally,sarkar2005dynamic} and Stochastic Block Models (SBM) \cite{matias2017statistical,ludkin2018dynamic,fu2009dynamic,jiang2020autoregressive,pensky2019dynamic}. \gdr{T}he study of \cite{suveges2022networks} 
uses a logistic autoregressive model 
\gr{utilising}
a community structure on the edges rather than the nodes of the graph.
 Two main differences of our work to the aforementioned studies \gdr{are} that (i) most of the studies focus on unweighted graphs, while we consider  weights on the edges of the network \gr{which carry the signal}, and (ii) we assume a fixed network structure with only the weights varying across time.

In the computer science literature, related work includes research on
traffic congestion in transportation, through the use of systems 
such as Global Positioning Systems which can collect data
from vehicles traversing road segments; a review can be found for example in
\cite{abdi2021review}. In this context, networks arise as road networks with nodes representing road junctions and edges representing road segments. In \cite{menelaou2018effective}, the authors propose two alternative approaches for predicting the travel times of vehicles for road segments (edges), using an Exponential Moving Average and a Multiple Linear Regression predictor. The key differences of our approach to the one proposed in \cite{menelaou2018effective} are that (i) we propose an autoregressive model (ii) \cite{menelaou2018effective} consider only direct neighbours to a segment while we allow nodes which are more than distance  1 away; we call these {\it higher-stage neighbours} and (iii) we further propose a technique for sparsifying the underlying network using lead-lag analysis of the time series. 

To model traffic flow it is natural to incorporate a spatial component in the model. \cite{min2011real} use spatial correlations among locations to model traffic flow time series. \cite{salamanis2015managing} extends the spatio-temporal autoregressive model by providing an alternative for the spatial correlation component, which now does not necessarily have a geographical interpretation, but rather includes information about neighbouring nodes that are detected using a Breadth First Search (BFS) algorithm. 
This study exhibits two main differences to our approach; (i) \cite{salamanis2015managing} model nodal time series, thus, the resulting approach is more related to the approaches proposed by \cite{zhu2017network} and \cite{knight2019generalised}; and (ii)
the proposed model includes only lags up to size 2 and neighbouring nodes up to stage 10, formulated specifically for the application of interest, and thus does not provide a general framework for network time series.

\paragraph{Main contribution.} In this paper, we provide a general framework for time series \gr{modelling and} prediction for networks with fixed structure and time-varying edge weights, \mc{fully leveraging the topology of the network} through identifying and incorporating information about neighbouring linkages into the model. Our approach is flexible as it can be implemented on any type of network, with no restriction on the specification of the lags and neighbour stages. Furthermore, our approach has good statistical properties while allowing for uncertainty quantification through confidence intervals.

\paragraph{Paper structure.} 
The rest of the paper is organised as follows. In Section \ref{sec2}, we describe the real data set  that has motivated the proposed model. In Section \ref{sec3}, we provide the necessary background to our approach. Section \ref{sec4} introduces the problem setup and the GNAR-edge model. In Section \ref{sec5}, we perform a range of synthetic data experiments to validate the GNAR-edge model performance in parameter estimation and prediction, also under model misspecification, and present its efficacy against baseline models. In Section \ref{sec6}, we fit the GNAR-edge model on the real data, and introduce a framework for network sparsification that exploits time series information. \gr{This framework uses scores from a lead-lag analysis to threshold edges.} We conclude with discussion and future work in Section \ref{sec7}. \gdr{The Supplementary material contains additional details of the  experiments }and real data results.

Code and synthetic data experiments are available at 

\url{https://github.com/mantziou/GNAR-edge-model}.

\section{A motivating data example}\label{sec2}

This work is motivated by an anonymised and aggregated UK business payments dataset made available to the Office for National Statistics by Pay.UK and Vocalink,  respectively the operator of and infrastructure provider to the UK’s retail interbank payment systems. Individual firm-to-firm transactions are aggregated to
\gdr{two-digit} SIC codes, as per latest list updated by Companies House on 9 January 2018 \cite{sic}, using a method developed by \gdr{the Office for National Statistics}. \gdr{Excluding {\it Activities of extraterritorial organisations and bodies} and {\it Dormant Company} there are 90 two-digit SIC codes.}  
The resulting dataset consists of multiple time series of monthly transactions between two-digit SIC codes 
\gdr{for a recording period starting from} January 2015 
\gdr{and ending in} July 2022, \gdr{having thus 91 time stamps}. 

This data \gdr{can be associated with} 
an underlying network structure, imposed by the pairs of transactions between SIC codes \gdr{as nodes} that comprise an edge list. \gdr{The presence, although not the weight, of an edge is taken to be constant acrosss time. We} record an edge \gdr{between two SIC codes} when there are no more than 85 zero payments observed in the time series for the recording period\gdr{, thus allowing for at most 6 of the 91 months to have a no payment recorded between the two SIC codes}. 
The resulting network has 90 nodes representing SIC codes, and 7,152 directed edges \gdr{(out of 8100 possible edges)} representing transactions. Transactions can occur within the same industry, hence, we observe self-loops in the network. The size\gr{s} of the transactions observed across time are 
represented as time-varying weights on the edges of the network. Thus there are two key assumptions under this construction: first, the network has a fixed structure, and second, the network has time-varying edge weights. 
On two-digit SIC aggregation level, the resulting time series comprise of mostly non-zero transactions. 
\gdr{One could instead have constructed a different network for each month, but there would be only very few}  
 changes in the network structure with respect to edges appearing or vanishing across time, \gdr{while adding considerable complexity to the model}. In light of this \gdr{feature of the data}, the assumption of a fixed network structure is natural for our problem.

Transactions \gdr{between SIC codes} 
can reflect the economic conditions in a country \cite{buda2022national}. For example, big disruptions can have an effect on the economy, and subsequently on transactions occurring between firms \cite{carvalho2021tracking}. The ability to accurately predict future transactions 
\gr{could} 
play a key role in identifying, \gr{navigating,}  and potentially \gr{even} preventing the occurrence of such 
\gdr{effects}. In addition, accounting for the underlying network formed by the associations between 
economic entities \gr{may} 
assist in capturing the propagation of shocks.
\gdr{For this type of data set, t}he research questions  
for the observed multivariate time series \gdr{which we address in this paper} are as follows.
\begin{itemize}
    \item Can we predict the  
    \gr{sizes} of future transactions accurately \gr{and with theoretical guarantees}?
    \item Can we leverage the information encoded by the underlying network structure to improve the forecasting task?
\end{itemize}

In Section \ref{sec4}, we introduce the GNAR-edge model that allows one to perform prediction under the assumption that a transaction occurring at time $t$ between two SIC codes is influenced not only by past transactions between the same two SIC codes but also by past values of neighbour transactions. \gdr{First we recall the general VAR model and outline its special case  introduced by \cite{knight2019generalised} that utilises a network structure.}

\section{Background}\label{sec3}

In the multiple time series setting, we observe time series of some fixed length $T$ for different variables $i=1,\ldots,K$. Let $X_i^t$ denote the value of variable $i$ at time $t=1,\ldots,T$. The VAR(L) model is an autoregressive model with maximum lag $L$, that has the following linear form,
\begin{equation}
    \label{eq:var} X_i^t=v_i+\alpha_{i_{1,1}}X_1^{t-1}+\alpha_{i_{2,1}}X_2^{t-1}+\ldots+\alpha_{i_{K,1}}X_K^{t-1}+\ldots+\alpha_{i_{1,L}}X_1^{t-L}+\ldots+\alpha_{i_{K,L}}X_K^{t-L}+u_i^t,\quad i=1,\ldots,K
\end{equation}
with intercept $v_i$, coefficients $\alpha_{i_{j,l}}, j=1, \ldots, K, l=1, \ldots, L$, and the innovations $(u_1^t, \ldots, u_K^t),$ $t=1, \ldots, T$, being
a $K-$dimensional white noise.
Equation \ref{eq:var} can be written compactly in vector form as 
\begin{equation} \label{eq:varform}
    \boldsymbol{X_t}=\boldsymbol{v}+\sum_{l=1}^L \boldsymbol{A_l} \boldsymbol{X_{t-l}}+\boldsymbol{U_t}
\end{equation}
with $\boldsymbol{X_t}=(X_1^t,\ldots,X_K^t)$,  $\boldsymbol{v}=(v_1,\ldots,v_K)$ denoting a 
$K-$dimensional vector of intercepts, $\boldsymbol{A_l}$ denoting a $(K\times K)$ coefficient matrix with $(i,j)$ entry $\alpha_{i_{j,l}}$ and $\boldsymbol{U_t}=(u_1^t, \ldots, u_K^t)$.
Under the VAR model, for each lag $l$ we regress on all variables $i=1,\ldots,K$; with fixed maximum lag $L$ there are $L K^2$ coefficients $a_{i_{j,l}}$ to estimate, which explains the high complexity of the VAR model which is $\mathcal{O}(K^2)$.  

In the case where a network structure is known for the variables, \gdr{to reduce the complexity of the model} \cite{knight2019generalised} propose a restricted case of the VAR model that utilises the neighbourhood structure of the network, \gdr{as follows}.
Let $\mathcal{G}=(V,E)$ denote a graph where $V= \{1,\ldots,n\}$ represents the set of $n$ nodes, and $E$ represents the set of edges with $E=\{(i,j)|i\rightarrow j; i,j\in V\}$ for a directed graph, or $E=\{(i,j)|i\leftrightarrow j; i,j\in V\}$ for an undirected graph. \gdr{For simplicity here we restrict attention to directed graphs; the adaptation to undirected graphs is straightforward.} 
In 
\cite{knight2019generalised}, 
a graph describes relationships among the set of \gdr{node} variables for which the time series are observed, thus 
$X_{i}^t$ represents the value of node $i$ at time $t$. \gr{They} 
assume that future values of a time series observed on node $i$ depend \gdr{not only} on past values of that node, but also on past values of times series observed on \gr{nodes which are} neighbours \gr{of} 
node $i$; 
the 
\gr{set} of 1-stage neighbour nodes of 
node $i$ \gr{is defined} as $\mathcal{N}^{(1)}(i)=\{j\in V/\{i\}:i\rightarrow j\}$ 
\gr{and for $r \ge 2$ the set of} 
$r$-stage neighbours of $i$ 
\gr{is defined as} $$\mathcal{N}^{(r)}(i)=\mathcal{N}\{\mathcal{N}^{(r-1)}(i)\}/[\{ \cup_{q=1}^{r-1}\mathcal{N}^{(q)}(i)\} \cup \{i\} ].$$ 
\gdr{The model allows for edge or node covariates
$c\in\{1,\ldots,C\}$, for some  $C\in \mathbb{N}$, which can be viewed as 
node or edge types.}
\gr{\cite{knight2019generalised} then} introduce a generalised autoregressive model with maximum lag $L$ and maximum stage $R_{l}$ for lag $l\in\{1,\ldots,L\}$ as follows,
\begin{equation}\label{gen_gnar}
X^t_{i} = \sum_{l=1}^{L} \Bigg(\alpha_{i,l}X^{t-l}_{i}+\sum_{c=1}^{C}\sum_{r=1}^{R_l}\beta_{l,r,c}\sum_{q\in \mathcal{N}_{t}^{r}(i)} w_{i,q,c}X^{t-l}_{q}\Bigg)+u_{i}^t, 
\end{equation}
with 
$w_{i,q,c}$ representing connection weights that
depend on the size of the neighbour set \gdr{and which could encode} 
edge-weight information in the case of weighted graphs. 
\gr{As t}he authors \gr{only} provide a framework for a static graph across time, 
despite the general form provided in \eqref{gen_gnar}, the $r$-stage neighbour set $\mathcal{N}^{r}(i)$ is not indexed by time.

In our work, we 
\gdr{take a similar approach as}
\cite{knight2019generalised} 
\gdr{but consider instead} the case where time series are observed on the edges of a fixed graph, rather than on the nodes. 
\gdr{This model requires} 
the notion of $r$-stage neighbouring edges for \gr{an} edge $\{i,j\}$,
\gr{which is introduced in the next section.}

\section{A network autoregressive model for edge time series}\label{sec4}

\subsection{Problem set-up and notation}\label{sec41}

Motivated by the real data application, our setting considers directed networks (graphs) on $n$ nodes with possible self-loops; we denote the adjacency matrix of graph $\mathcal{G}$ by $A$. We let $N = n^2$ denote the number of possible  edges and $|E|=K$ the number of edges in the static network. 
Our model assumes that $\mathcal{G}$ is fixed, but that there is a time varying process on $E$ which can be viewed as a matrix-valued process $V_t, t \ge 0$ of non-negative weights. We denote by  $\boldsymbol{X^t}=A\odot V_t$ the time-varying weighted adjacency matrix of the network time series at time $t$, with $\odot$ denoting the element-wise product, or Hadamard product, between matrices $A$ and $V_t$. Thus, $X^t_{ij}$ represents the weight of edge $\{i,j\}$ at time $t$.

\begin{figure}[!h]
    \centering
    \includegraphics[scale=.5]{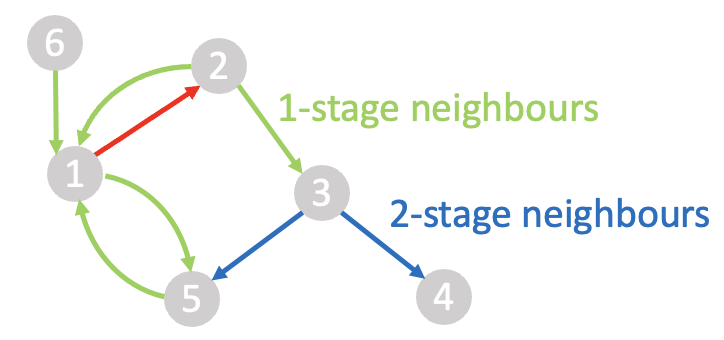}
    \caption{\gr{An} example of 1-stage and 2-stage neighbours for edge $1\rightarrow 2$.}
    \label{neighbor_example}
\end{figure}

\gr{We define the set of} 1-stage neighbouring edges for some edge $\{i,j\}$ 
\gr{as the set of all} 
edges \gr{which are} incident to \gr{at least one of the} nodes $i,j$; \gr{formally,} 
$$\mathcal{N}^{1}(\{i,j\})=\{\{k,l\}\in \delta^{+}(i)\cup \delta^{-}(i)\cup \delta^{+}(j)\cup \delta^{-}(j): \{k,l\}\neq \{i,j\} \}$$ with $\delta^{+}(\cdot),\delta^{-}(\cdot)$ denoting the sets of outgoing and incoming edges  of a node, respectively. \gr{For $r \ge 2$} 
\gdr{we define}
the set of $r$-stage neighbouring edges of the edge $\{i,j\}$ 
\gdr{as} the set $$\mathcal{N}^{r}(\{i,j\})=\mathcal{N}\{\mathcal{N}^{r-1}(\{i,j\})\}/[\{\cup_{q=1}^{r-1}\mathcal{N}^{q}(\{i,j\})\}\cup \{\{i,j\} \}].$$
\gr{As the network is fixed, these \gdr{sets} 
do not depend on time.} 
 Figure \ref{neighbor_example} \gr{shows} 
 an example of the 1-stage and 2-stage neighbours for the edge $1\rightarrow 2$ \gdr{in a toy graph}. 

\subsection{The GNAR-edge model}\label{sec42}

Inspired by \cite{knight2019generalised}, we assume that the weight $X^t_{ij}$ of an edge between nodes $i,j$ at time $t$ depends not only on its past values, but also on past values of its neighbouring edges. 
Thus, the GNAR-edge model has the following general form for maximum lag $L$
\begin{equation}\label{gen_gnar_edge}
X^t_{ij} = \sum_{l=1}^{L} \Bigg(\alpha_{ij,l}X^{t-l}_{ij}+\sum_{r=1}^{R_l}\beta_{l,r}\sum_{m,n:\{m,n\}\in \mathcal{N}^{r}(\{i,j\})} w_{ij,mn}X^{t-l}_{mn}\Bigg)+u_{ij}^t, 
\end{equation}
with $\alpha_{ij,l}$ denoting the standard autoregressive parameters at lag $l$ for edge $\{i,j\}$, $\beta_{l,r}$ denoting the parameters for the effect of $r$-stage neighbouring edges at lag $l$, $w_{ij,mn}=|N^r(\{i,j\})|^{-1}$ denoting the normalising weight for $X_{mn}^{t-l}$ which equally weights all neighbouring edges of edge $\{i,j\}$ and the innovations $u_{ij}^t$ denoting
white noise with mean 0 and variance $\sigma^2$. 
The model in equation \ref{gen_gnar_edge} is abbreviated as GNAR-edge($L$,[$R_1$,\ldots,$R_L$]) where the first argument $L$  denotes the maximum lag and each element of the second argument $R_l$ denotes the maximum stage neighbours for lag $l$.

We note that the weights $w_{ij,mn}$ do not necessarily need to reflect the size of the neighbour set but could perhaps reflect associations between edge time series measured by some correlation function, or by some lead-lag metric \citep{bennett2022lead}. While in our model, we are not assuming any edge or node covariate information, this setting is amenable to a straightforward extension.

Equation \ref{gen_gnar_edge} can be further expressed in vector format as follows. 
We assume a labelling  $l : E\rightarrow \{1,\ldots,K\}$ on the set of edges.
Let $\boldsymbol{X_t}=(X_{1}^t,\ldots,X_{K}^t)$ 
denote the vector of weights of the $K$ labelled edges, at time $t$. Let $\boldsymbol{A_l}=\text{diag}\{\alpha_{1,l},\ldots,\alpha_{K,l}\}$ denote the $K\times K$ matrix enclosing the autoregressive parameters for lag $l$, using the relabelling of the indices ${ij}, \{i,j\} \in E$, and let
\begin{displaymath}
\boldsymbol{W^{(r)}}=
\begin{pmatrix}
    w_{11}\mathbb{I}\{1\in\mathcal{N}^{(r)}(1)\} & \dots  & w_{1K}\mathbb{I}\{K\in\mathcal{N}^{(r)}(1)\} \\
    \vdots  & \ddots & \vdots \\
    w_{K1}\mathbb{I}\{1\in\mathcal{N}^{(r)}(K)\}  & \dots  & w_{KK}\mathbb{I}\{K\in\mathcal{N}^{(r)}(K)\}
\end{pmatrix}
\end{displaymath}
denote the $K\times K$ matrix enclosing the normalising weights of the neighbouring edges, for edges $1,\ldots,K$. Thence, equation \ref{gen_gnar_edge} can be written in vector form as
\begin{equation}\label{gen_gnar_edge_vec}
    \boldsymbol{X_t}=\sum_{l=1}^L (\boldsymbol{A_l}\boldsymbol{X_{t-l}}+\sum_{r=1}^{R_l} \beta_{l,r} \boldsymbol{W^{(r)}}\boldsymbol{X_{t-l}})+\boldsymbol{U_t},
\end{equation}
with $\boldsymbol{U_t}=(u_1^t,\ldots,u_K^t)$. Equation \ref{gen_gnar_edge_vec} can be further reduced to 
\begin{equation}\label{gen_gnar_edge_vec_2}
    \boldsymbol{X_t}=\sum_{l=1}^L \boldsymbol{\Psi_l} \boldsymbol{X_{t-l}}+\boldsymbol{U_t}
\end{equation}
with $\boldsymbol{\Psi_l}=\boldsymbol{A_l}+\sum_{r=1}^{R_l} \beta_{l,r} \boldsymbol{W^{(r)}}$, for $t=L+1,\ldots,T$; this is of the VAR($L$) form given in equation \ref{eq:varform}.
Thus, as a special case of a VAR process, equation \ref{gen_gnar_edge_vec_2} can be written as a linear model in matrix form, 
\begin{equation}\label{matform}
    \boldsymbol{X}=\boldsymbol{B}\boldsymbol{Z}+\boldsymbol{U}
\end{equation}
where $\boldsymbol{B}=[\boldsymbol{\Psi_1},\ldots,\boldsymbol{\Psi_L}]$, 
$\boldsymbol{X}=[X_{L+1},\ldots,X_{T}]$, $\boldsymbol{Z}=[Z_L,\ldots,Z_{T-1}]$ with $Z_t=[X_{t},\ldots,X_{t-L+1}]$ and $\boldsymbol{U}=[\boldsymbol{U_{L+1}},\ldots,\boldsymbol{U_{T}}]$. Hence, the OLS estimators are given in closed form by
\begin{equation*}
    \hat{\boldsymbol{B}}=(\boldsymbol{Z}^T \boldsymbol{Z})^{-1}\boldsymbol{Z}^T \boldsymbol{X},
\end{equation*}
assuming that $\boldsymbol{Z}^T \boldsymbol{Z}$ is invertible. 
The covariance of the parameter estimates is $\text{cov}[\hat{\boldsymbol{B}}]=\Sigma_u\otimes(\boldsymbol{Z}^T \boldsymbol{Z})^{-1}$, with $\otimes$ denoting the Kronecker product and $\Sigma_u$ denoting the covariance matrix of the white noise innovations \citep{tsay2013multivariate}. In practice, $\Sigma_u$ is not known thus a consistent estimator of $\Sigma_u$ is given by $\hat{\Sigma}_u=T^{-1}(\boldsymbol{X}-\hat{\boldsymbol{B}}\boldsymbol{Z})(\boldsymbol{X}-\hat{\boldsymbol{B}}\boldsymbol{Z})^{T}$ \citep{knight2019generalised,lutkepohl2005new}, resulting in $\hat{\text{cov}}[\hat{\boldsymbol{B}}]=\hat{\Sigma}_u\otimes(\boldsymbol{Z}^T \boldsymbol{Z})^{-1}$.

Similarly to the GNAR model \citep{knight2019generalised}, a sufficient condition for the GNAR-edge model \eqref{gen_gnar_edge} to be stationary is 
\begin{equation*}
    \max_{\{i,j\}}\sum_{l=1}^L\left(|\alpha_{ij,l}|+\sum_{r=1}^{R_l}|\beta_{l,r}|\right)<1 
\end{equation*}

The parameters encoding the network effect $\beta_{l,r}$ are not edge-specific, which is beneficial when dealing with large networks. On the other hand, the autoregressive parameters $\alpha_{ij,l}$ are edge-specific which can increase the model complexity for large networks. Thus, we further consider the restricted case of the GNAR-edge model in which a global $\alpha_{l}$ for each lag size $l$ is assumed; \gr{we call this the {\it global GNAR-edge model}}. In the rest of this paper, we consider the global GNAR-edge \gr{model}.

\section{Synthetic data experiments}\label{sec5}

In this section, we perform synthetic data experiments to explore the performance of the GNAR-edge model in parameter estimation and prediction. In our experiments, we consider both moderate\gr{ly} sized networks as well as larger networks with similar size to that of the real data.

\subsection{Estimation performance}\label{sec51}
\subsubsection{Moderately sized networks}\label{sec511}
\gdr{Here w}e investigate the performance of \gr{the} GNAR-edge model in estimating the model coefficients for five different parameter specifications and three different network structures.

We first explore the case of a simple network structure,  namely an Erd\"{o}s-R\'{e}nyi (ER) network model. 
We generate 50 ER 20-node directed networks with 168 edges each, and simulate time series on the edges of each ER network using the GNAR-edge model, for different parameter specifications as detailed in Table \ref{sim_reg}, \gdr{using the notation from Section \ref{sec42}}. 
The length of the simulated time series is 200. To simulate time series from the GNAR-edge model, we initialise using the \gdr{multivariate} Standard Normal distribution. For each simulation regime, we generate 50 different seeded data sets and fit the GNAR-edge model to each data set. Indicatively, in Appendix \ref{AppA1}, Figure \ref{er_sbm_rdp} shows one of the 50 simulated ER networks with density 0.44 and diameter 3. \gdr{Any neighbourhood stage which is at least as large as the diameter would yield an autoregressive model in which each edge is modelled using all other edges in the network}. \gdr{Hence our experiments include specifications of only up to}
2-stage neighbouring edges.

\begin{table}[htb!]
\centering
\begin{tabular}{clcc}
\multicolumn{1}{l}{} &                                   & \textbf{$\boldsymbol{\alpha}$} & \textbf{$\boldsymbol{\beta}$}                    \\ \cline{2-4} 
1                    & \textbf{GNAR-edge(1, {[}1{]})}     & 0.2            & 0.3                              \\
2                    & \textbf{GNAR-edge(1, {[}2{]})}     & 0.2            & (0.3,0.4)                        \\
3                    & \textbf{GNAR-edge(3, {[}1,1,1{]})} & (0.2,0.4,-0.6) & (0.2,0.1,-0.2)                   \\
4                    & \textbf{GNAR-edge(3, {[}2,2,2{]})} & (0.2,0.4,-0.6) & ((0.3,0.1),(0.1,0.1),(-0.2,0.3)) \\
5                    & \textbf{GNAR-edge(3, {[}2,0,0{]})} & (0.2,0.4,-0.6) & ((0.3,0.4),0,0)          \\   \cline{2-4}     
\end{tabular}
\caption{Simulation regimes for moderate\gr{ly} sized networks. 
}
\label{sim_reg}
\end{table}
As Simulation Regime 4 is the most complex of these regimes, we report the results from fitting the GNAR-edge model on the different seeded data sets in this regime in Table \ref{covsim4}; the results for the other regimes can be found in the Supplementary Material.
Specifically, for each model parameter, we report the frequency of the 95\% confidence intervals enclosing the true value, as well as the Root Mean Square Error (RMSE) for the estimated coefficients with respect to their true values. We observe a very good performance of the GNAR-edge model in estimating the true model parameters, as indicated by the very small, close to 0 RMSE obtained for all parameters. We further notice that for the parameters capturing the network effect $\boldsymbol{\beta}$ the RMSE is slightly bigger, indicating that the inference task might be harder for those parameters. Similarly, the very good estimation performance of the GNAR-edge model is observed from the high coverage rates for the 95\% confidence intervals which are all close to nominal level. \gdr{Section 1 in the} 
Supplementary Material
further present tables with results for the \gdr{other} 
simulation regimes, for which we observe similar performance.

\begin{table}[htb!]
\begin{center}
\begin{tabular}{ccccccc}
         & \multicolumn{6}{c}{Regime 4}                        \\ \cline{2-7} 
         & \multicolumn{2}{c}{ER} & \multicolumn{2}{c}{SBM} & \multicolumn{2}{c}{RDP} \\ \cline{2-7} 
         & \begin{tabular}[c]{@{}c@{}}Coverage\\ (\%)\end{tabular}    & RMSE     & \begin{tabular}[c]{@{}c@{}}Coverage\\ (\%)\end{tabular}     & RMSE     & \begin{tabular}[c]{@{}c@{}}Coverage\\ (\%)\end{tabular}     & RMSE     \\ \hline
$\alpha_1$ & 100         & 0.004    & 0.98         & 0.003    & 0.94         & 0.004    \\
$\beta_{1,1}$ & 0.98        & 0.02     & 0.94         & 0.02     & 0.94         & 0.02     \\
$\beta_{1,2}$ & 0.94        & 0.05     & 0.94         & 0.05     & 0.96         & 0.04     \\
$\alpha_2$ & 0.98        & 0.003    & 0.94         & 0.004    & 0.98         & 0.004    \\
$\beta_{2,1}$ & 0.96        & 0.02     & 0.96         & 0.02     & 0.92         & 0.02     \\
$\beta_{2,2}$ & 0.94        & 0.06     & 0.92         & 0.05     & 0.92         & 0.06     \\
$\alpha_3$ & 0.98        & 0.004    & 0.96         & 0.004    & 0.92         & 0.005    \\
$\beta_{3,1}$ & 0.92        & 0.02     & 0.96         & 0.02     & 0.98         & 0.02     \\
$\beta_{3,2}$ & 0.98        & 0.04     & 0.98         & 0.05     & 0.98         & 0.05     \\ \hline
\end{tabular}
\caption{Coverage rates for $95\%$ confidence interval enclosing the true value and RMSE for estimated coefficients, across 50 replications,  for simulation regime 4.}
\label{covsim4}
\end{center}
\end{table}

Real networks typically exhibit a community structure. A widely-used model for networks with nodes forming underlying communities is the Stochastic Block Model (SBM). We consider a directed, 20-node SBM model with $K=2$ blocks \gdr{and} edge-formation probabilities \gdr{which are} \gr{larger} for nodes within the same community 
\gdr{than} 
to nodes from different communities. Specifically, we set the probability of an edge to occur between nodes in the same block to $p_{11}=p_{22}=0.7$, and the probability of connection between nodes in different blocks to $p_{12}=0.1$ and $p_{21}=0.2$, respectively, so that the resulting graphs have density approximately 0.4,  
\gdr{similar to  the fixed density of the} ER model\gdr{s}. 
In Appendix \ref{AppA1}, Figure \ref{er_sbm_rdp} shows one of the simulated SBM networks with density 0.44 and diameter 3. We simulate time series on the edges of the SBM networks using the GNAR-edge model and the different parameter specifications as presented in Table \ref{sim_reg}. We replicate the experiments 50 times and  present the results from fitting the GNAR-edge model to the simulated data in Table \ref{covsim4}, and Tables 1-4 in \gdr{the} Supplementary Material, Section 1. Similarly to the ER case, we observe that the estimated coefficients are very close to their true values as indicated by the small RMSE, as well as the high coverage rates close to nominal level, for all simulation regimes, except for $\beta_{11}$ in \gdr{S}imulation \gdr{R}egime 2, which is slightly smaller \gdr{(0.88)}.

In addition to the ER and SBM \gdr{models} 
we 
consider a Random Geometric Graph structure. The random geometric graphs we consider in our analysis are Random Dot Product (RDP) graphs, which assume that nodes have a latent position in space, and the probability of an edge between two nodes is given by the dot product of their latent positions. Specifically, we sample two-dimensional vectors representing positions of the nodes on the surface of a sphere with radius 0.7, such that the resulting networks have density approximately 0.44, similarly to \gr{our} ER and SBM networks. 
In Appendix \ref{AppA1}, Figure \ref{er_sbm_rdp} shows 
 one of the simulated RDP networks using a layout that respects \gr{the positions of the nodes} 
using the shortest path matrix \cite{igraph}. The diameter of the RDP graph in Figure \ref{er_sbm_rdp} is 3. 
We \gr{then} generate time series on the edges of the RDP networks according to \gdr{the}  simulation regimes in Table \ref{sim_reg}, and present the results for 50 replications of our experiments, in Table \ref{covsim4}, and Tables 1-4 in \gdr{the} Supplementary \gdr{M}aterial, Section 1. We observe an overall similar performance of the model for the RDP network, with estimated coefficients close to their true values, and high frequency of 95\% confidence intervals enclosing the true values.

\gr{Concluding, we find that the parameter estimation 
\gdr{performs} well on these networks, and that there is not a substantial difference between the performance of the estimation for the different underlying networks.} In the next section, we further explore the estimation accuracy of our model for network sizes \gr{which are} similar to the real data.

\subsubsection{\gr{A} large  network}\label{sec512}

In this subsection we explore the performance of the GNAR-edge model for \gr{simulated networks} 
\gr{which are} similar in size to that of the real data application. In \gdr{the} real data application, we observe a very densely connected graph, with network density approximately equal to 0.9.  \gdr{F}rom our simulations on moderately sized networks, 
we do not expect different network structures to have a \gr{substantial} impact on the model performance for the case of densely connected graphs. In light of this, \gdr{from now on we focus on} an Erd\"{o}s-R\'{e}nyi graph \gdr{as it} 
has a \gdr{very} simple structure. Specifically, we generate an Erd\"{o}s-R\'{e}nyi graph with $|V|=86$ nodes and $|E|=6858$ edges, and simulate time series on its edges with 90 time stamps, under 
\gdr{two} simulation regimes, presented in Table \ref{sim_reg2}. 

\begin{table}[htb!]
\centering
\begin{tabular}{clcc}
\multicolumn{1}{l}{} &                                   & {$\boldsymbol{\alpha}$} & {$\boldsymbol{\beta}$}                    \\ \cline{2-4} 
1                    & \textbf{GNAR-edge(4,{[}1,1,1,1{]})} & (-0.6,-0.4,-0.2,-0.1) & (0.2,0.1,0.3,0.05)                   \\
2                    & \textbf{GNAR-edge(4,{[}2,2,2,2{]})} & (-0.6,-0.4,-0.2,-0.1) & ((0.4,-0.4),(0.3,-0.4),(0.5,-0.3),(0.05,-0.1))  
\end{tabular}
\caption{Simulation regimes for the large ER network.}
\label{sim_reg2}
\end{table}

\begin{table}[htb!]
\centering
\begin{minipage}[c]{0.4\textwidth}
\begin{tabular}{lcc}
       & \multicolumn{2}{c}{\textbf{GNAR-edge(4,{[}1,1,1,1{]})}}                                           \\ \cline{2-3} 
       & \textbf{Estimated} & \textbf{\begin{tabular}[c]{@{}c@{}}95\% confidence \\ interval\end{tabular}} \\ \hline
$\alpha_1$ & -0.59              & (-0.60 -0.59)                                                                \\
$\alpha_2$ & -0.4               & (-0.40,-0.39)                                                                \\
$\alpha_3$ & -0.19              & (-0.20,-0.19)                                                                \\
$\alpha_4$ & -0.1               & (-0.10,-0.09)                                                                \\
$\beta_{1,1}$ & 0.19               & (0.14,0.23)                                                                  \\
$\beta_{2,1}$ & 0.11               & (0.06,0.16)                                                                  \\
$\beta_{3,1}$ & 0.33               & (0.28,0.38)                                                                  \\
$\beta_{4,1}$ & 0.05               & (0.005,0.09)                                                                 \\ \hline
\end{tabular}
\caption{Simulation results for \gr{a} large network, under Simulation Regime 1.}
\label{sim_res2}
\end{minipage}
\hspace{2cm}
\begin{minipage}[c]{0.4\textwidth}
\centering
\begin{tabular}{lcc}
       & \multicolumn{2}{c}{\textbf{GNAR-edge(4,{[}2,2,2,2{]})}}                                           \\ \cline{2-3} 
       & \textbf{Estimated} & \textbf{\begin{tabular}[c]{@{}c@{}}95\% confidence \\ interval\end{tabular}} \\ \hline
$\alpha_1$ & -0.6             & (-0.605 -0.6007)                                                                \\
$\alpha_2$ & -0.4               & (-0.40,-0.39)                                                                \\
$\alpha_3$ & -0.2              & (-0.20,-0.19)                                                                \\
$\alpha_4$ & -0.1               & (-0.10,-0.09)                                                                \\
$\beta_{1,1}$ & 0.38               & (0.33,0.43)                                                                  \\
$\beta_{1,2}$ & -0.43               & (-0.63,-0.22)                                                                  \\
$\beta_{2,1}$ & 0.24               & (0.19,0.29)                                                                  \\
$\beta_{2,2}$ & -0.51               & (-0.75,-0.27) 
                                             \\
$\beta_{3,1}$ & 0.48               & (0.43,0.53)                                                                  \\
$\beta_{3,2}$ & -0.41               & (-0.65,-0.17)                                                                  \\
$\beta_{4,1}$ & 0.049               & (0.003,0.09)                                                                  \\
$\beta_{4,2}$ & -0.13               & (-0.33,0.07)                                                                 \\ \hline
\end{tabular}
\caption{Simulation results for  \gr{a} large network, under Simulation Regime 2.}
\label{sim_res22}
\end{minipage}
\end{table}
Tables \ref{sim_res2} and \ref{sim_res22}, show the estimated model coefficients along with their confidence intervals, for each simulation regime.
We observe that in all simulation scenarios, the estimated coefficients are close to their true values, and the 95\% confidence intervals enclose the real values. The results give credence that the GNAR-edge model has good estimation performance in settings similar to that of the real data.

\subsection{Predictive performance}\label{sec52}
We now assess the performance of the GNAR-edge model in predicting future values of the edge time series. We compare the predictive performance of the GNAR-edge \gdr{model} to the \gdr{VAR model \eqref{eq:var}} 
that does not 
explicitly account for network structure, and \gdr{to fitting an AR model} 
for each time series individually \gdr{and ignoring dependence between time series}. 
\gr{As in the} 
previous section, experiments are performed for networks of both moderate and large sizes.

\subsubsection{Moderately sized networks}\label{sec521}

\begin{figure}[htb!]
    \centering
    \includegraphics[scale=.55]{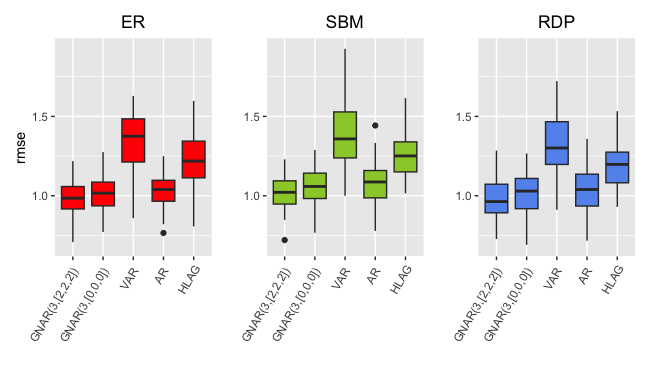}
    \vspace{-7mm}
    \caption{Distribution of RMSE for various time series models ($x$ axis) 
    and network structures (colour), for networks with density approximately 0.1.}
    \label{rmse_sim_rep}
\end{figure}

\begin{figure}[htb!]
    \centering
    \includegraphics[scale=.55]{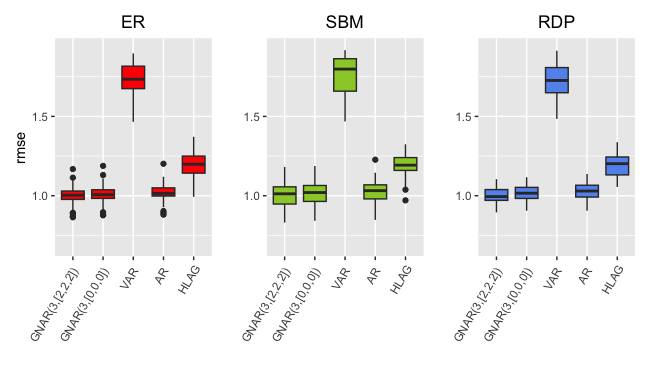}
    \vspace{-7mm}
    \caption{Distribution of RMSE for various time series models ($x$ axis) and network structures (colour), for networks with density approximately 0.4.}
    \label{rmse_sim_rep2}
\end{figure}

We first consider networks with $|V|=20$ nodes and  
\gr{different} network structures. 
The baseline models we consider for comparisons to our approach are the VAR and AR models. These are two widely used models for analysing multivariate 
time series. A main restriction of the VAR model is its increasing complexity as the number of time series increases. In light of this, the implementation of the VAR model on very densely connected networks (such as the data size in our application) can quickly become infeasible.
To reduce dimensionality of the VAR model, we fit it on synthetic data in two ways. First, we use the \texttt{restrict} function which sets  any coefficient to zero whose absolute t-statistic value is less than two, similarly to \cite{knight2019generalised}. Second, we use the R package \texttt{BigVAR} to implement
the Lasso-type VAR estimation process (HLAG) approach by \cite{nicholson2020high}.
In our experiments for moderate\gr{ly} sized networks, we consider two 
\gr{different} densities of the graphs,  
\gr{namely} 0.1 and 0.4.
For both
network densities, we use the simulation regime 4 (Table \ref{sim_reg}) to simulate time series on the edges of the networks using a GNAR-edge(3,[2,2,2]) model. We perform repetitions of this simulation experiment, by simulating multiple networks and time series on the edges with the same parameter specifications, for each network structure considered. In addition to the VAR and AR models, we consider a GNAR-edge model assuming no neighbour structure, to compare results to the case where neighbours are included, as per simulation regime used to generate the time series. We fit each model on the simulated time series excluding the last time stamp, and predict the last time stamp using the fitted model. To evaluate the predictive performance we use the Root Mean Square Error (RMSE). The results of the predictive performance of each model, for multiple repetitions are presented in Figures \ref{rmse_sim_rep} for graphs with density approximately 0.1, and Figure \ref{rmse_sim_rep2} 
for graphs with density approximately 0.4.


The results show that the 
\gdr{GNAR-edge(3,[2,2,2])} model, \gdr{which includes}  
neighbour structure,  outperforms all alternative models considered, under all scenarios. In addition, we notice that the performance does not change significantly among the different network structures assumed. 
The VAR model fitted with the \texttt{restrict} function (VAR) has the worst performance for both networks, with density 0.1 and 0.4. For networks with density 0.1, we are able to fit the VAR model with lag 3, while for networks with density 0.4 this is infeasible due to model complexity; thus, we fit a VAR with lag 1. This \gr{simpler model choice may} 
explain the poorer performance of the VAR model for the case of  network density 0.4 \gr{compared to 0.1}. Using the HLAG approach by \cite{nicholson2020high}, 
we are able to fit the VAR model with lag 3 for both networks with density 0.1 and 0.4. We observe that the HLAG estimation
yields better predictions compared to the simple VAR case. However, HLAG does not outperform the GNAR-edge model with neighbour structure. Moreover, the AR model which is fitted with lag 3 for all scenarios 
outperforms the VAR fitted with the HLAG approach in all simulation regimes.
Fitting the GNAR-edge model with and without neighbour structure does not lead to  
considerably different results; however, 
 for the case of  density 0.1, the better performance of the GNAR-edge model with neighbour structure is more evident. 
 
 Driven by these results, we further explore the effect of different network densities on the predictive performance of the GNAR-edge model with and without neighbour structure. 
Specifically, for each network structure considered, we simulate 20-node networks with densities ranging from 0.1 to 0.4. For each network density, we simulate 50 graphs and for each simulated graph we further simulate time series of weights for its edges for the same simulation regime considered above. For each resulting network time series, we fit the \gdr{models} GNAR-edge(3,[2,2,2]) and GNAR-edge(3,[0,0,0]) and evaluate 
\gdr{their} predictive performance by 
the RMSE. In Appendix \ref{AppA2}, Figures \ref{rmse_sim}, \ref{rmse_sim2} and \ref{rmse_sim3} show the distribution of the RMSE for each network density and network structure. We observe \gr{from Figures \ref{rmse_sim}, \ref{rmse_sim2} and \ref{rmse_sim3}}  that for different network densities, the distribution of the RMSE varies. Notably, we observe an overall better predictive performance of the GNAR-edge model with neighbour structure for smaller densities. \gr{As} 
the density increases, the performance of the two \gdr{GNAR-edge} models becomes more similar, \gdr{echoing our findings from}  
Figures \ref{rmse_sim_rep} and \ref{rmse_sim_rep2}.  
Nonetheless, in no scenario \gr{does} the GNAR-edge model with no neighbours outperform the GNAR-edge model with neighbours. These results indicate that sparser networks can be more informative for prediction, but also that different network structures have an effect in the performance, as indicated by the variance of the RMSE for sparser networks.

 This finding has triggered our interest in developing a technique for sparsifying the network resulting from the real data, \mc{as a potential form of regularization}, to possibly assist in predictive performance, as presented in Section \ref{sec6}. 
 
 It is also worth noting here that it is anticipated not to see a drastic change in RMSE between the neighbour and no-neighbour assumption, as the data are simulated under the 2-stage neighbour assumption. 
 \gdr{I}ncreasing the neighbour stages \gdr{c}ould increase the difference in RMSE between the neighbour and no-neighbour structure, as we 
 see in \gdr{the} next subsection.
\subsubsection{Large networks}\label{sec522}

Next we explore the predictive performance of the model\gdr{s} for data that are of similar size to the real data example. First, we consider networks with 86 nodes and network density 0.1 and simulate edge time series with 90 time stamps, for the three  
\gr{different} network structures (SBM, ER, RDP). The simulation regime used to generate time series on the edges of the networks is Simulation Regime 1 from Table \ref{sim_reg2}. We repeat each experiment 50 times. In this simulation study, we do not compare to the VAR model as it had poorest performance for moderately sized networks (as seen in Section \ref{sec521}) even when using the dimensionality reduction approach HLAG.
In Appendix \ref{AppA3}, Figure \ref{rmse_sim_large} shows the distribution of the RMSE from fitting the  
\gr{different} models, for 50 repetitions, on networks with density approximately 0.1. The diameter of the simulated networks is approximately 4. To further mimic the real data sizes, we set up the same experiments for networks with density 0.9, which is \gdr{close to} the network density observed for the real data. As networks with density close to 0.9 are almost complete networks, we perform the experiments only for the ER model. In Appendix \ref{AppA3}, Figure \ref{rmse_sim_large2} shows the results for networks with density approximately 0.9 and diameter 2. 

\begin{figure}[htb!]
\centering
\begin{subfigure}{.48\textwidth}
    \centering
    \includegraphics[scale=.42]{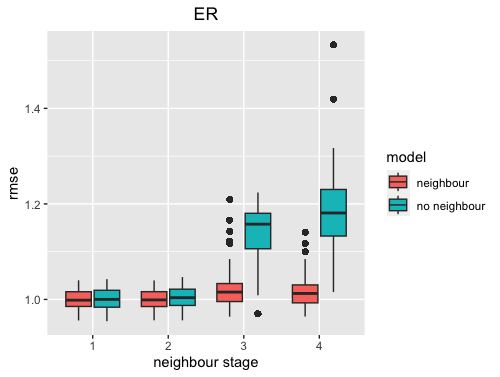}
    \end{subfigure}
\begin{subfigure}{.48\textwidth}
    \centering
    \includegraphics[scale=.42]{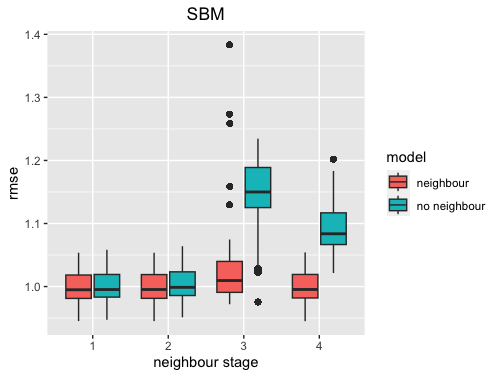}
    \end{subfigure}
    \caption{RMSE distribution for GNAR-edge \gdr{models} with and without neighbour structure, for different stages ($x$ axis), \gdr{on networks with density approximately 0.1. Left: Erd\"{o}s-R\'{e}nyi networks; right: Stochastic block model networks.}}
    \label{rmse_sim_stag_er_sbm}
\end{figure}

For this large networks case we obtain similar results to the results obtained from simulations for moderately-sized networks in Section \ref{sec521}, however, we notice an even poorer performance of the AR model compared to the GNAR-edge model. 
This can possibly be explained by the increased number of time series to be estimated, which under the AR model are considered individually while the GNAR-edge model analyses them jointly. What is also worth noting is the similar performance of the GNAR-edge model\gdr{s}  with \gdr{and without} neighbour structure,  
especially  
noticeable for the network density 0.9. This can be explained by the simulation regime specified in this experiment, where only 1-stage neighbours are assumed. To illustrate that, we compare the performance of the GNAR-edge model with and without neighbour structure for a range of higher stages. Figure \ref{rmse_sim_stag_er_sbm} shows the distribution of the RMSE for 50 repetitions of simulation regimes with 1, 2, 3 and 4 stage neighbours. As we do not observe significant changes in performance across the network models considered, we perform these experiments for the ER and SBM structures only. 
The results suggest that the GNAR-edge model with neighbour structure \gdr{overall performs better than} 
models with no neighbour structure for 
networks with a more complex topology, and higher stage neighbours.

\subsection{Model misspecification}\label{sec53}

In this section, we investigate the performance of the GNAR-edge model in estimation and prediction, under model misspecification scenarios. Specifically, we consider three alternative types of model misspecification. First, we explore the estimation and predictive performance of the model for simulated network time series with heavy-tailed noise, and specifically we consider a t-distribution with 3 and 10 degrees of freedom (chosen to ensure that the distribution has finite variance). Second, we examine the estimation performance of the GNAR-edge model in settings where innovations are correlated across time. Third, we explore the model performance under misspecification of the network connectivity patterns. For each case, we consider both moderately-sized and large networks.

\subsubsection{Moderately-sized networks}\label{sec531}

\begin{figure}[htb!]
    \centering
    \includegraphics[scale=.45]{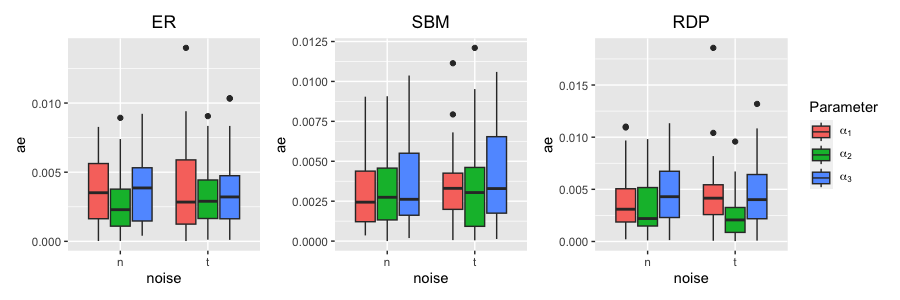}
    \caption{Distribution of absolute error for autoregressive parameters $\alpha_1,\alpha_2,\alpha_3$, for different network structures, and  normally distributed (n) and t-distributed with 3 degrees of freedom (t) innovations.}
    \label{est_ae_alphas}
\end{figure}

For moderately sized networks, we consider the simulated data of Section \ref{sec511}, and specifically the $4^{th}$ simulation regime presented in Table \ref{sim_reg} as this is the most complex regime with respect to the number of parameters to be estimated. We simulate  20-node networks and time series on the edges of the networks under the GNAR-edge model with parameters as per simulation regime 4 in Table \ref{sim_reg}, while the innovations are simulated from a heavy-tailed distribution. Notably, we consider a t-distribution with two different parameter specifications that control the probability mass in the tails of the distribution. First we consider a more heavy-tailed case assuming 3 degrees of freedom, and second we consider a less heavy-tailed case assuming 10 degrees of freedom. We perform 50 repetitions of each experiment and report the absolute error between the estimated coefficients and their true values. Figure \ref{est_ae_alphas} shows the distribution of the absolute error for all autoregressive parameters, under normally distributed (n) and t-distributed with 3 degrees of freedom (t) innovations, for different network structures. The results for the stage-1 and stage-2 neighbour effect parameters are similar to the results presented for the autoregressive parameters, thus they are presented in Appendix \ref{AppA4}, Figures \ref{est_ae_betas1} and \ref{est_ae_betas2}.

\begin{figure}[htb!]
    \centering
    \includegraphics[scale=.5]{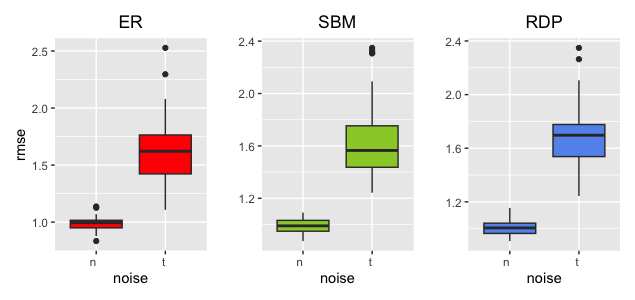}
    \caption{Distribution of RMSE for normally distributed (n) and t-distributed with 3 degrees of freedom (t) innovations, for different network structures.}
    \label{rmse_noise}
\end{figure}

The estimation performance of the GNAR-edge model on simulated data with t-distributed innovations is similar to that with normally distributed 
innovations. However, we notice that for the autoregressive parameters, the distribution of the absolute error has more outliers for the heavy-tailed case compared to the normally distributed innovations. Despite the overall similar estimation performance for the two different innovation distributions, we anticipate that the prediction performance will be poorer for the synthetic data with heavy-tailed innovations. We investigate that by evaluating the predictive performance of the GNAR-edge model on the synthetic data with normally distributed and heavy-tailed innovations, for the same simulation regime. To do that we fit the GNAR-edge model to the network time series excluding the last time stamp, use the fitted model to predict the last time stamp and evaluate the prediction performance using the RMSE. The boxplots in Figure \ref{rmse_noise} show the distribution of the RMSE for 50 repetitions of the experiment. As expected, the RMSE obtained for the heavier-tailed network time series is bigger compared to the case of normally distributed errors. In Appendix \ref{AppA4} we also present the results from experiments for t-distributed innovations with 10 degrees of freedom, for which we observe a similar estimation performance to that of 3 degrees of freedom, and a slightly better predictive performance as the innovations are less heavy-tailed for degrees of freedom 10.

\begin{figure}[htb!]
    \centering
    \includegraphics[scale=.42]{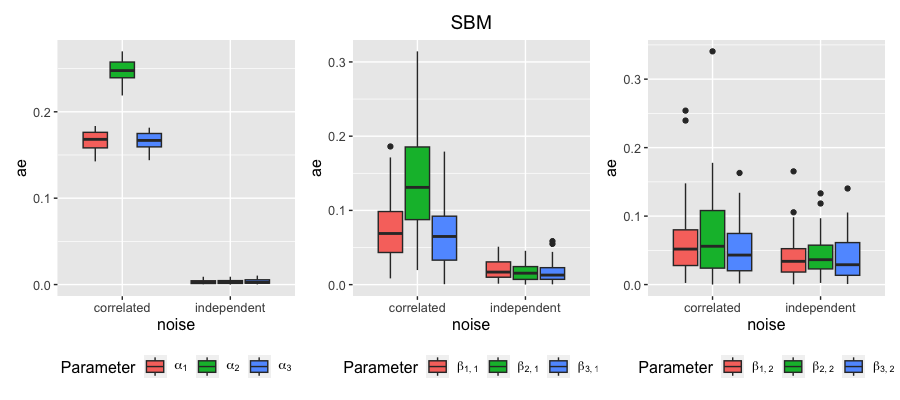}
    \caption{Distribution of absolute error for all model parameters for settings with independent and correlated innovations with correlation 0.5, for SBM network structure.}
    \label{est_ae_all_corr_sbm}
\end{figure}

We now examine the model performance in settings where the innovations of the network time series are correlated across time. In this simulation scenario, we simulate network time series with innovations coming from a multivariate   normal distribution with mean 0 and variance 1, for two different correlation sizes among all pairs of normally distributed variables equal to 0.1 and 0.5, for less and more strongly correlated innovations respectively. Similarly to previous experiments, we consider the $4^{th}$ simulation regime to generate the network time series under the GNAR-edge model with correlated errors, and fit the GNAR-edge model to the simulated data, for 50 repetitions. First we present the results for correlation equal to 0.5. In Figure \ref{est_ae_all_corr_sbm},  we present the distribution of the absolute error for the estimated model parameters across the 50 repetitions for the SBM network structure, for correlated and independent innovations. Similar results are obtained for the ER and RDP network structures, thus they are presented in Appendix \ref{AppA4}, Figures \ref{est_ae_allpar_corr_er} and \ref{est_ae_allpar_corr_rdp}.

The estimation
performance of the GNAR-edge model changes to a different extent for the different model parameters. 
For the autoregressive parameters, in Figure \ref{est_ae_all_corr_sbm}  we see a clear effect of the correlated errors on the estimation performance against the independent errors case, with the estimation performance for the former being considerably poorer.
The colours are not distinguishable for the autoregressive parameters in the independent errors setting in Figure \ref{est_ae_all_corr_sbm}, due to the large difference in the scale of the absolute error in the correlated errors setting.) For a closer look to the distributions of the absolute error for the autoregressive parameters in the independent noise setting see Figure \ref{est_ae_alphas}, middle plot, (n) case. For the 1-stage neighbour effect parameters, we see again that the estimation performance of the GNAR-edge is poorer for the correlated innovations case, however, the scale of the absolute error is now smaller compared to the scale of the absolute error for the autoregressive parameters. Similarly for the 2-stage neighbour effect, we notice an even smaller scale of the absolute error for the correlated innovations, however the estimation performance is still poorer compared to the independent innovations. This is a reasonable finding given that the dependence of the errors is introduced across time, rather than across the edges of the network, thus having a stronger effect on the autoregressive parameters compared to the network effect parameters. Overall, we observe that the estimation performance of the GNAR-edge model is poorer for all model parameters when innovations are correlated across time, as anticipated.

We repeat the above experiments for correlation equal to 0.1 and present the results in Appendix \ref{AppA4}. We observe a  
similar performance of the GNAR-edge model for 0.1 correlation to that of 0.5 correlation among innovations; however, the scale of the absolute error in this setting is smaller, due to the smaller correlation introduced.

\begin{figure}[htb!]
    \centering
    \includegraphics[scale=.47]{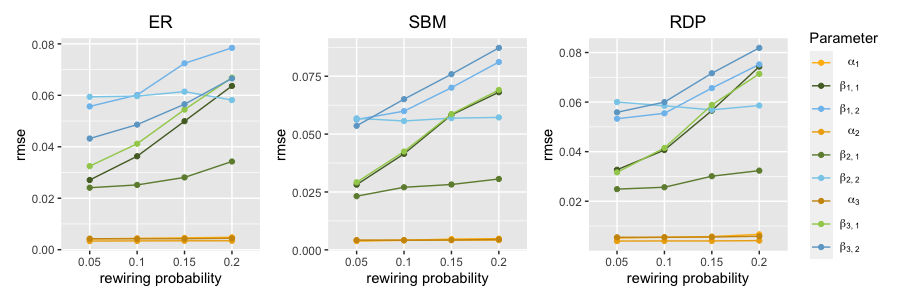}
    \caption{RMSE of estimated model parameters after 50 repetitions, and various perturbations of network structure (x axis).}
    \label{rmse_est_rewiring}
\end{figure}

\medskip 
We also investigate the estimation performance of the GNAR-edge model under misspecification of the neighbourhood structure, for the same simulation regime, as follows. 
After simulating the network time series, we perform edge rewiring for different rewiring probabilities using the \texttt{rewire} function from the R package \texttt{igraph}, and fit the GNAR-edge model to the data using the rewired network. We specify the rewiring method \texttt{each$\_$edge} within the \texttt{rewire} function that rewires the endpoints of edges with a constant probability uniformly at random to a new node in the graph. In our experiments, we consider rewiring probabilities of 0.05, 0.1, 0.15 and 0.2. Each of these rewiring probabilities correspond to an approximate change of 25, 45, 65 and 80 edges, respectively, in the original network with 380 possible edges. We perform 50 repetitions and report the RMSE of the estimated coefficients with respect to their true values. Figure \ref{rmse_est_rewiring} shows how the RMSE scales for increasing rewiring probabilities. The mean Hamming distance between the rewired networks and corresponding true networks among the 50 repetitions is 0.067, 0.12, 0.17 and 0.21, respectively for each rewiring probability 0.05, 0.1, 0.15 and 0.2, indicatively for the ER network structure. Similar sizes of mean Hamming distance are obtained for the SBM and RDP networks. 

The RMSE for the network effect parameters increases for increasing perturbations of the network structure, as it would be anticipated. Specifically, we observe a more drastic increase of the RMSE for the network effect parameters for lag $L=1,3$, while
for the autoregressive parameters and the the network effect parameters for lag $L=2$, we observe a more stable RMSE. This is especially anticipated for the autoregressive parameters as they are not capturing network information.

\subsubsection{Large networks}\label{sec532}

In this section, we repeat the same simulation experiments for our model misspecification settings, for large networks. Specifically, we consider 86-node networks and simulate network time series under simulation regime 1 of Table \ref{sim_reg2}. Similarly to moderately-sized network in the previous section, we first explore the estimation performance of the GNAR-edge model for heavy-tailed innovations. For the large networks setting, we consider only the case of t-distribution with 3 degrees of freedom as this is the heavier-tailed case, and for moderately-sized networks we did not observe considerably different results between 3 and 10 degrees of freedom. The results of the simulation experiments are presented in Appendix \ref{AppA5}, in Figures \ref{est_ae_alphas_large}, \ref{est_ae_betas1_large} and \ref{rmse_noise_large}, which show the distribution of the absolute error of the estimated parameters and the RMSE of the predictions, under normally distributed and t-distributed innovations, for different network structures. Similarly to results of the previous section, we notice that the estimation performance of the GNAR-edge model is similar in both settings of correlated and independent innovations, 
while, as seen in Figure \ref{rmse_noise_large} the prediction performance of the GNAR-edge model is considerably worse in the case of heavy-tailed network time series, as anticipated.

\begin{figure}[htb!]
    \centering
    \includegraphics[scale=.4]{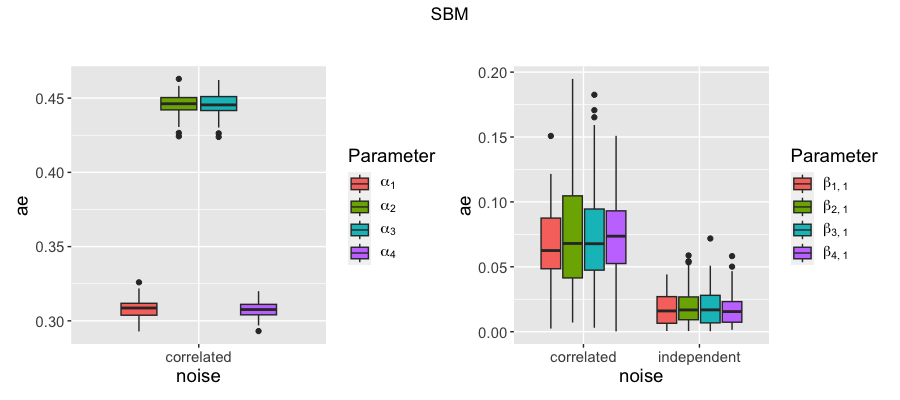}
    \caption{Distribution of absolute error for all model parameters for settings with independent and correlated innovations with correlation 0.5, for SBM network structure.}
    \label{est_ae_all_sbm_corr_large}
\end{figure}

\medskip 
We also examine the effect of time correlated innovations for large networks. Similarly to our experiments for moderately-sized networks, we simulate edge time series for large networks, under two scenarios, (i) with i.i.d.\,innovations simulated from a standard normal distribution and (ii) with innovations simulated from multivariate standard normal distribution with correlation 0.5 across variables. In this experiments we consider only the 0.5 correlation scenario, as it is the setting that had the greatest effect in the estimation performance for moderately-sized networks, as seen in Section \ref{sec531}. We perform again 50 repetitions and present the results for the estimation performance of the GNAR-edge model in Figure \ref{est_ae_all_sbm_corr_large} for the SBM network structure. The results for the ER and RDP networks are presented in Appendix \ref{AppA5}, Figures \ref{est_ae_all_er_corr_large} and \ref{est_ae_all_rdp_corr_large}.
Figure \ref{est_ae_all_sbm_corr_large}, left panel, shows only the distribution of the absolute error for the correlated errors setting as colours would not be distinguishable if results were included in the same plot for the independent errors setting, due to the difference in the scale of the absolute error. For the results in the independent error case see Figures \ref{est_ae_alphas_large}, \ref{est_ae_betas1_large} (n) case. Similarly to the moderately-sized networks, we observe that the estimation performance of the GNAR-edge model is considerably poorer for the correlated innovations setting, particularly notable 
for the autoregressive parameters of the model as seen in Figure \ref{est_ae_all_sbm_corr_large} left panel and Figure \ref{est_ae_alphas_large} middle panel (n) case.

\medskip 
We further explore the estimation performance of the GNAR-edge model under misspecification of the neighbourhood structure, in the large networks setting. We rewire the edges of the true network with rewiring probabilities 0.05, 0.1, 0.15 and 0.2, which correspond to an approximate change of 139, 270, 395 and 512 edges, respectively, in the original network with 7310 possible edges. We perform 50 repetitions and report the RMSE of the estimated parameters in Figure \ref{rmse_est_rewiring_large}. The mean Hamming distance between the rewired networks and corresponding true networks among the 50 repetitions is 0.018, 0.037, 0.054 and 0.07, respectively for each rewiring probability, indicatively for the ER network structure, with similar sizes of mean Hamming distance obtained for the SBM and RDP networks. 

Similarly to moderately-sized networks, we notice that the estimation of the autoregressive parameters is not affected by the perturbed networks, however, the estimated parameters for the network effect move further away from their true values as the rewiring probability increases. Nonetheless, for the network effect parameters at lag $l=2,4$, and especially for the network effect parameter at lag $l=4$, $\beta_{4,1}$, the effect of the network misspecification for increasing rewiring probability is relatively more stable.

\section{Real data application}\label{sec6}
\gr{In this section, w}e consider the multivariate time series \gr{data set described in Section \ref{sec2}} resulting from anonymised and aggregated transactions between 2-digit SIC codes, ranging from January 2015 to July 2022. The network emerging from the pairwise transactions comprises of 90 nodes (2-digit SIC) and 7,152 directed edges (transactions), being a very densely connected network with density approximately 0.9. \gdr{Our} 
simulations have \mcr{shown that} sparser networks can assist in better prediction performance. Thus,
\gdr{here} we propose a technique for sparsifying networks arising from multivariate time series.

\subsection{\mcr{Network sparsification}} 

\gdr{Our technique is based on} 
lead-lag analysis.
The investigation of leading and lagging relationships between time series is a common approach for analysing multivariate time series systems \cite{wu2010detecting}. 
A time series A is said to lead a time series B if past values of time series A are more strongly correlated with future values of time series B \gr{than with past values of time series B}. Thus, lead-lag analysis can be very informative for forecasting. To investigate such relationships between multiple time series, 
pairwise distances of the time series with respect to \gdr{a lead-lag metric can be employed. This results in a lead-lag matrix with rows and columns corresponding to \mcr{each individual} time series.} 
\cite{bennett2022lead} present a range of alternatives for specifying a lead-lag metric. In our study, we investigate the lead-lag relationships among the observed multiple time series of transactions, specifying Pearson's correlation and \mcr{considering} the difference of the cross-correlation function at lag $\in \left\{ -1,1\right\}$ (ccf-lag1 as presented in \cite{bennett2022lead}).

Using the lead-lag analysis, we sparsify the network by identifying the top leading edges. To \mcr{this end, we consider} the row sums of the lead-lag matrix, \gdr{which we take as indicator of} 
the size of leadingness of each time series over the rest of the time series. We consider the top 801 leading edges in the network, resulting in a network with density 0.1, which is the \mcr{setting} for which the smallest RMSE has been achieved in simulations. 

\gdr{As a check of the network choice, we assess whether the resulting network has small-world structure, a characteristic}
of many real networks;
\gdr{in particular,} \cite{de2016topologic} study an economic network of transactions in Estonia, and they observe that \mcr{it} is characterised by small-world \gr{behaviour}.  \gdr{T}he two network properties that indicate small-world \gr{behaviour} \gdr{are}
an average shortest path length \gr{of similar size as that of a Bernoulli random graph with same network density, (often) coupled with an average local}  clustering coefficient \gr{which is high compared to that of a corresponding Bernoulli random graph.}
\gdr{Our} resulting network has \gr{an average local}  clustering coefficient 0.2, which is \gr{considerably} larger than the expected clustering coefficient of 0.1 for a 90-node random graph with 0.1 density. However, the average shortest path length is 2.3, which is \gr{only} slightly higher than the expected average shortest path length of 2 for a random 90-node graph with 0.1 density. \mc{
\gdr{T}hese results \gdr{taken together} hint at the possible presence of a small-world structure \gdr{in our sparsified network.}}

\subsection{Predictive performance} 
\begin{figure}[htb!]
    \centering
    \includegraphics[scale=.7]{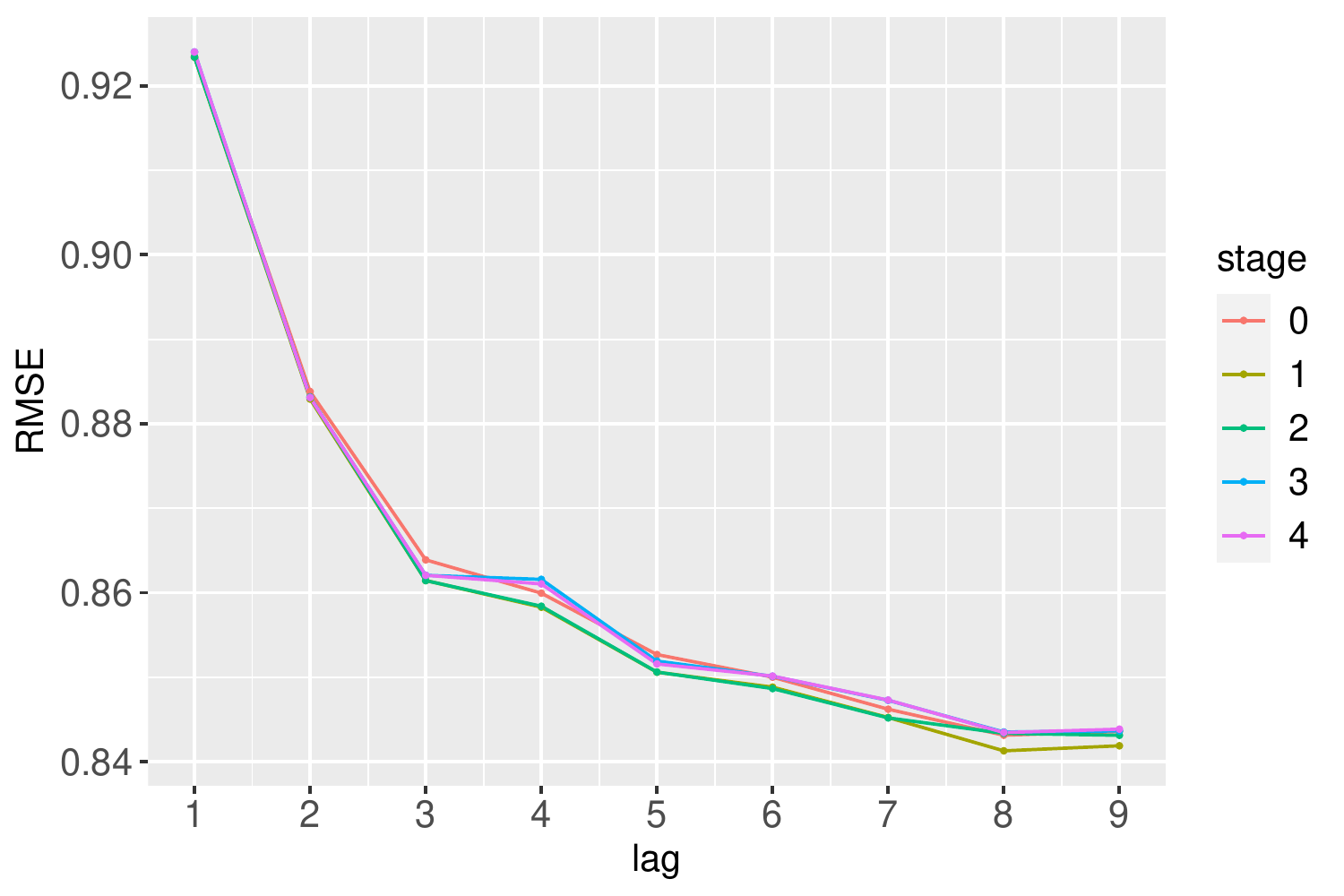}
    \caption{RMSE from predicting the last time stamp of the time series, for various lags ($x$ axis) and neighbour stages (colour).}
    \label{datarmse}
\end{figure}

After sparsifying the network, we consider as training sample the time series observed on the edges of the network up to June 2022. We pre-process the time series by  \gdr{carrying out} 
\mcr{first-order} differencing to remove trend, and \gdr{by} dividing by the standard deviation for each edge of the training sample \gdr{to reduce heterogeneity}. 
We fit the GNAR-edge model to the training data and
predict \gr{the} last time stamp,  
July 2022. \gdr{The} 
predictive performance \gdr{is again evaluated} using the
RMSE over all the time series. We examine the predictive performance for a range of lags $l=\{1,\ldots,9\}$ and neighbour stages $r=\{0,1,2,3,4\}$ and present the results in Figure \ref{datarmse}.

Figure \ref{datarmse} shows that increasing the lag size leads to better prediction performance, as \mcr{revealed by} the decreasing RMSE. Notably, we observe a steep decrease of the RMSE up to lag 3, \mcr{beyond which} the RMSE gradually decreases until it  
achieves a (possibly local) minimum at
lag 8. We further observe similar predictive performance up to and including lag 2 for all different neighbour stages considered for the GNAR-edge model, while after lag 2, the performance of the alternative models starts to vary. In addition, 1-stage and 2-stage neighbour  \gr{models} 
have 
\gr{similar} predictive performance across lags as seen by the overlapping lines. However, 
for lag 8 when \gr{the} RMSE \gr{is minimised}, \gr{the} 1-stage neighbour \gr{model}  has the best performance among the models, with the 
\gr{remaining} models having very similar RMSEs.

We compare these results to the results from fitting an AR model, \gr{the model class} which has shown the most competitive performance to our \gdr{GNAR-edge models} 
in our synthetic data experiments. 
 \mcr{The results in Figure \ref{datarmse} showcase} that the best performing model is the GNAR-edge model for lag 8 and 1-stage neighbours. 
In light of this, we fit 
\gr{an} AR(8) \gr{model} to the time series resulting after network sparsification and predict the last time stamp, which yields 
an RMSE of 0.9433. \gdr{This RMSE}  
is considerably higher  
\gr{than} the RMSE obtained from the GNAR-edge model with lag 8 and 1-stage neighbours,  being 0.8412. In \gdr{the} Supplementary Material, Section 2, we 
present the results from fitting an AR model for a range of lags from 1 up to 7,  \mcr{all of which attain} an RMSE very close to the RMSE obtained from fitting an AR(8) \gdr{model}. To further illuminate the utility of the sparsification step in improving the prediction performance of our model, we consider the whole network, and fit the GNAR-edge model with lag 8 and 1-stage neighbours.
\gr{T}he RMSE obtained when considering the whole network is 1.3864, which is considerably larger than the RMSE of 0.8412, obtained for the sparsified network.

\begin{figure}[htb!]
\centering
\begin{subfigure}{.45\textwidth}
    \centering
    \includegraphics[scale=.5]{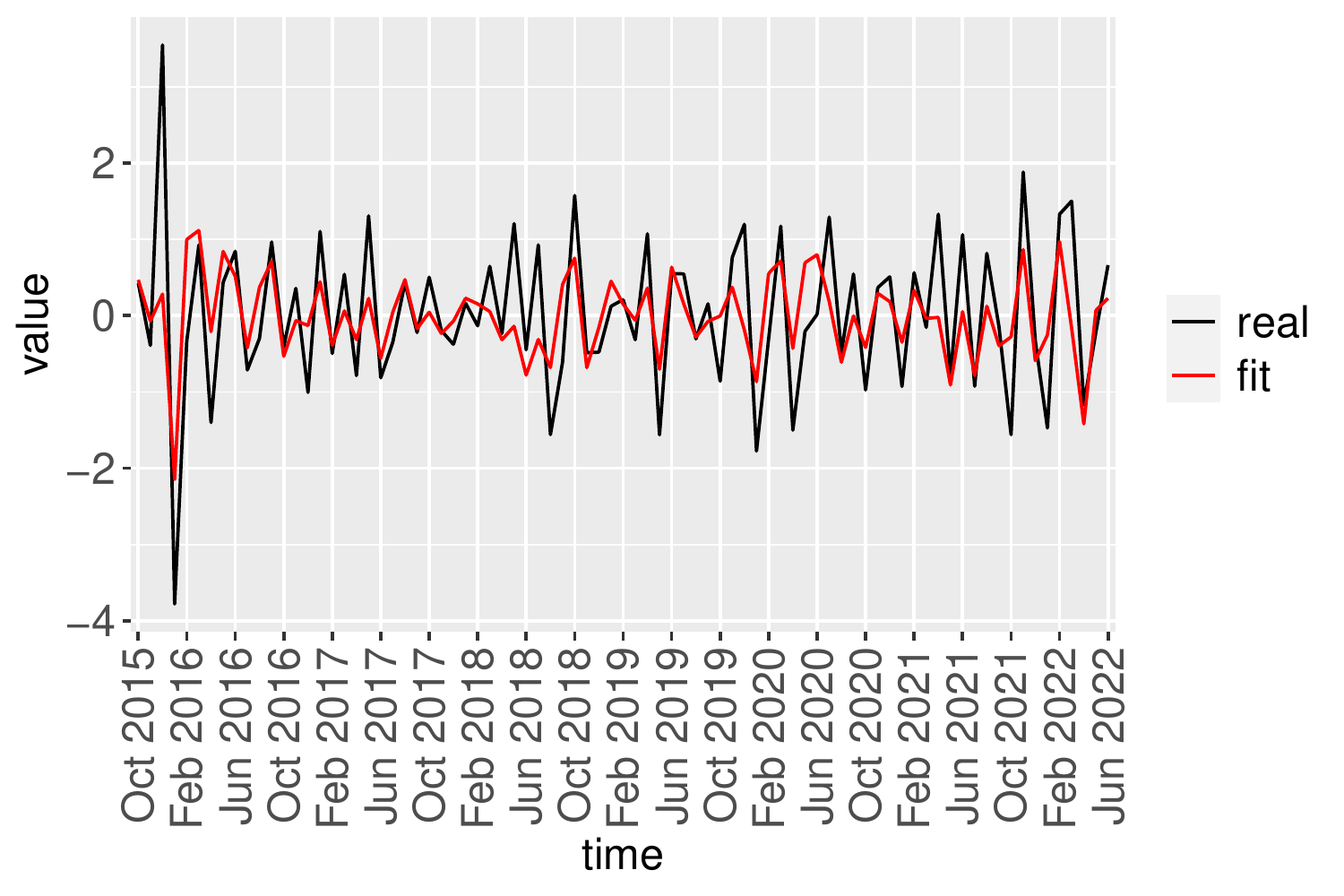}
    \end{subfigure}
\begin{subfigure}{.45\textwidth}
    \centering
    \includegraphics[scale=.5]{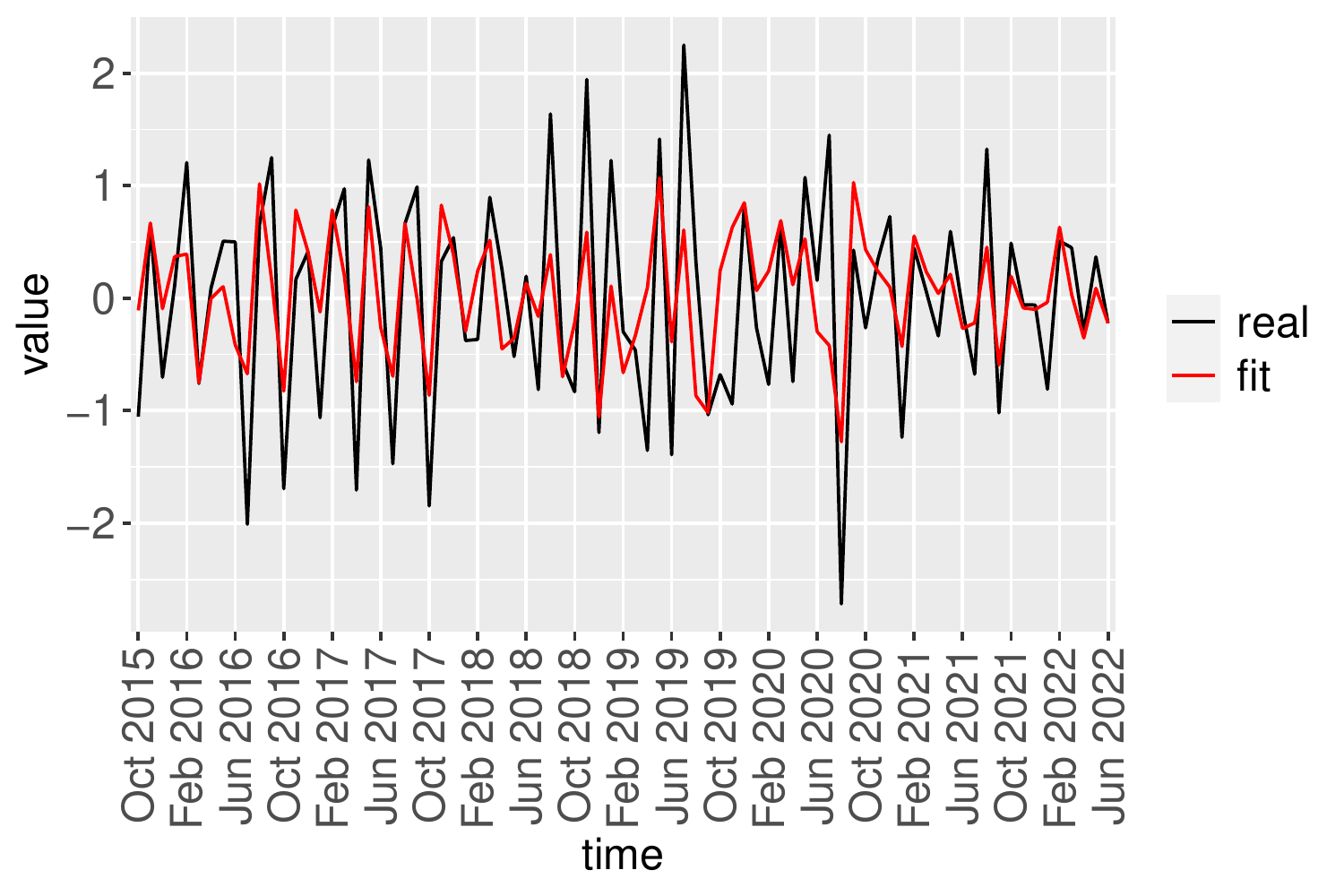}
    \end{subfigure} 
    \caption{Fitted lines (red) on \gdr{two} real time series (black) from fitting \gdr{a} GNAR-edge model  \gdr{with lag 8 and 1-stage neighbours} on \gdr{the} sparsified network.}
        \label{fitt_lines}
\end{figure}

From Figure \ref{datarmse}, we notice that for each lag, there is 
\gr{no considerable} variability in the scale of the RMSE across the different stages of neighbours. This can \gr{perhaps} be explained by the fact that the best fitting model as suggested by the results, is the model with 1-stage neighbours, and as seen in synthetic data experiments for 1-stage neighbours we do not observe 
\gr{considerable} changes in the RMSE versus the case with no neighbours assumed. 
Figure \ref{fitt_lines} illustrates the fit of the GNAR-edge model with lag 8 and 1-stage neighbours, indicatively for \gdr{two} of the 801 edge time series. We observe an overall very good fit of the GNAR-edge model on the real data; however, larger peaks of the time series are not fully captured by the model.

\begin{figure}[htb!]
    \centering
    \includegraphics[scale=.4]{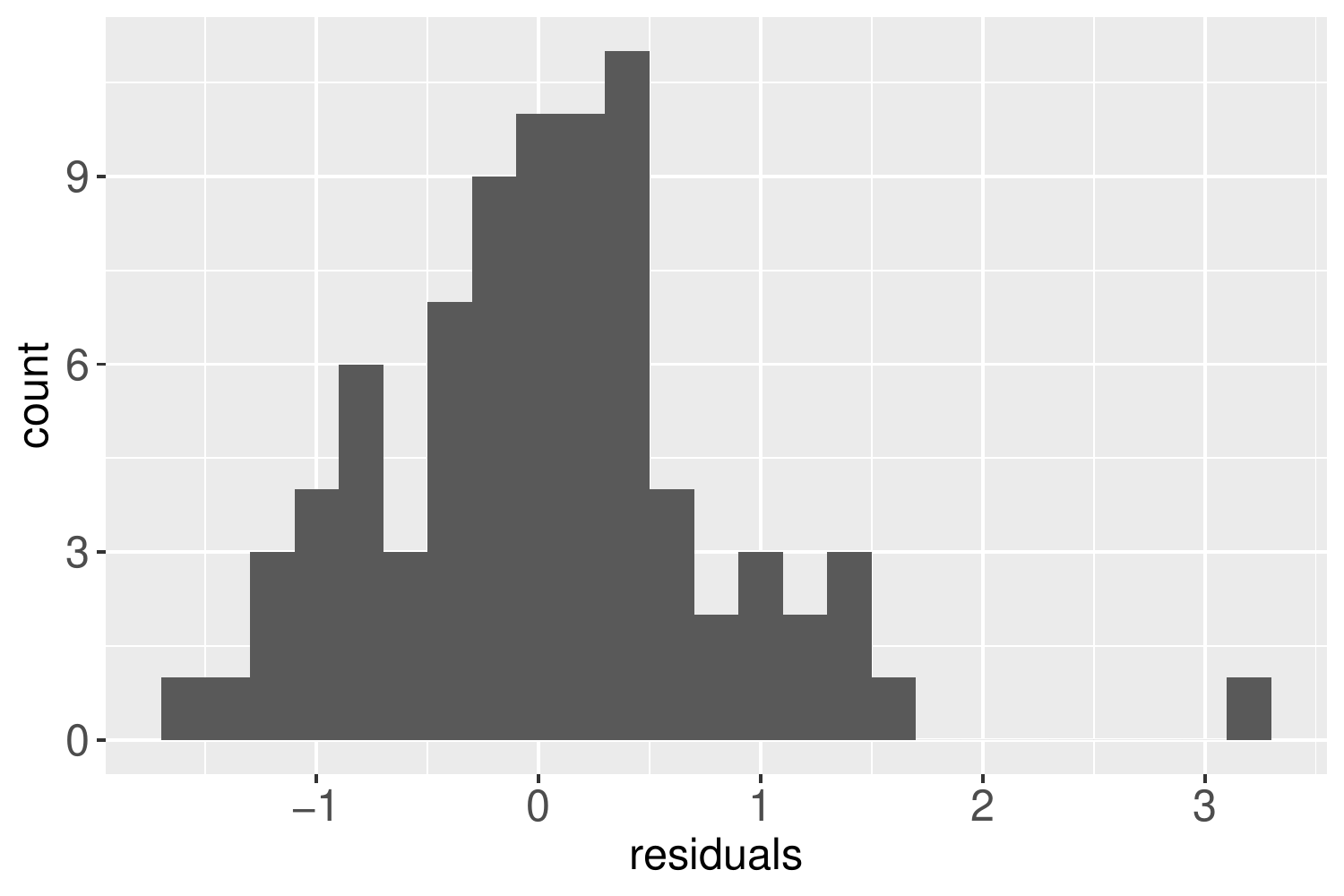}
    \qquad
    \centering
    \includegraphics[scale=.4]{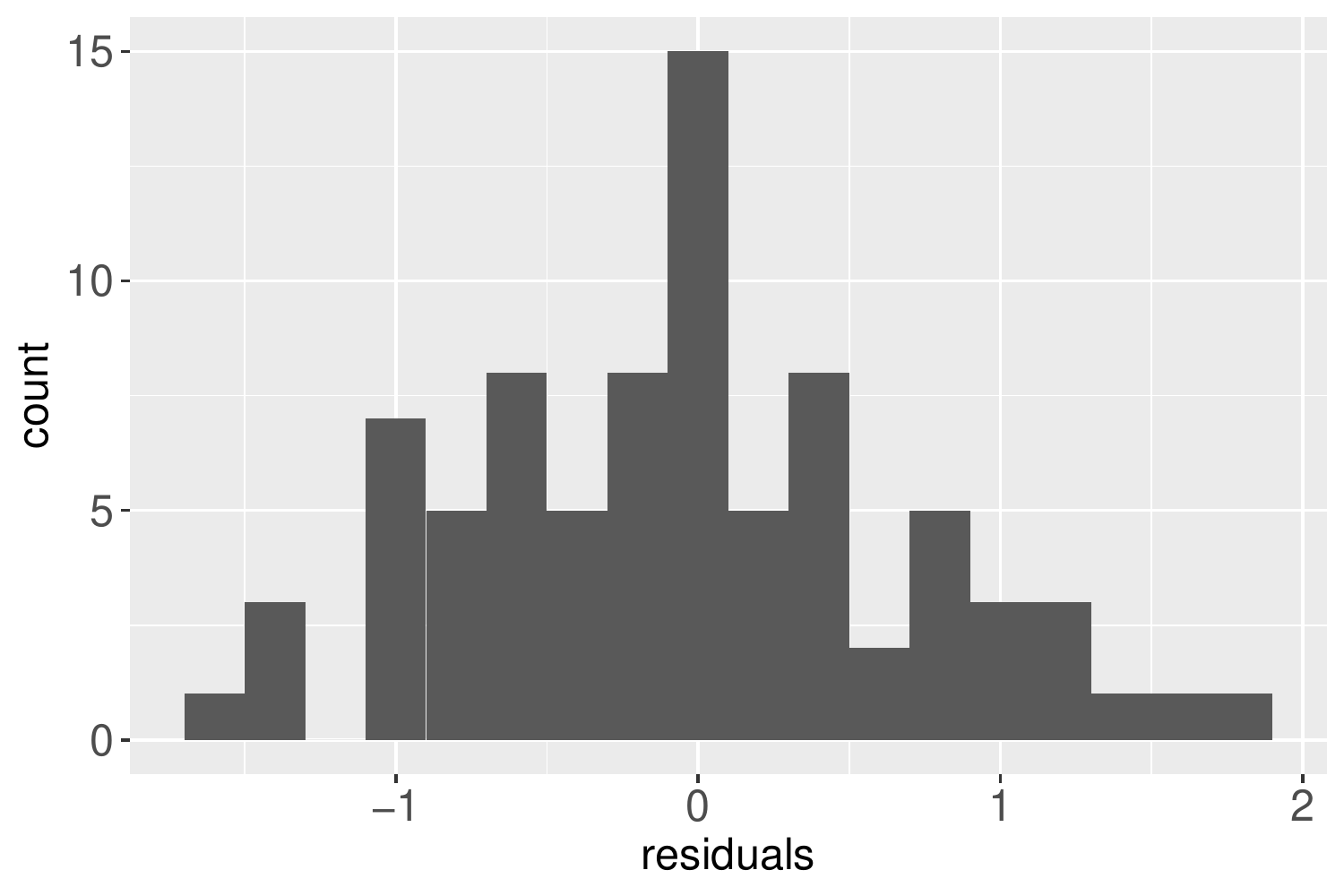}
\vfill
\centering
    \includegraphics[scale=.4]{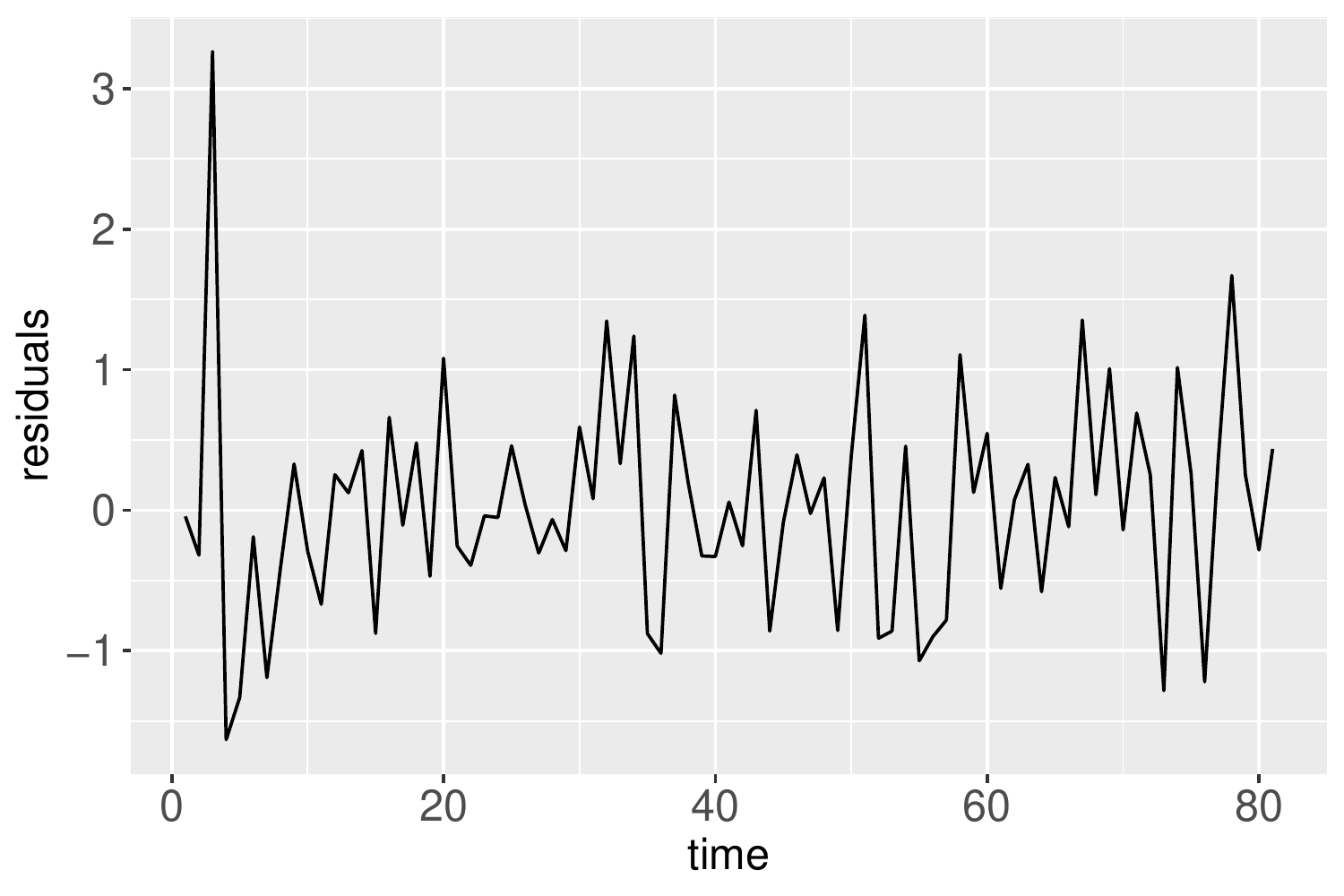}
    \qquad
    \centering
    \includegraphics[scale=.4]{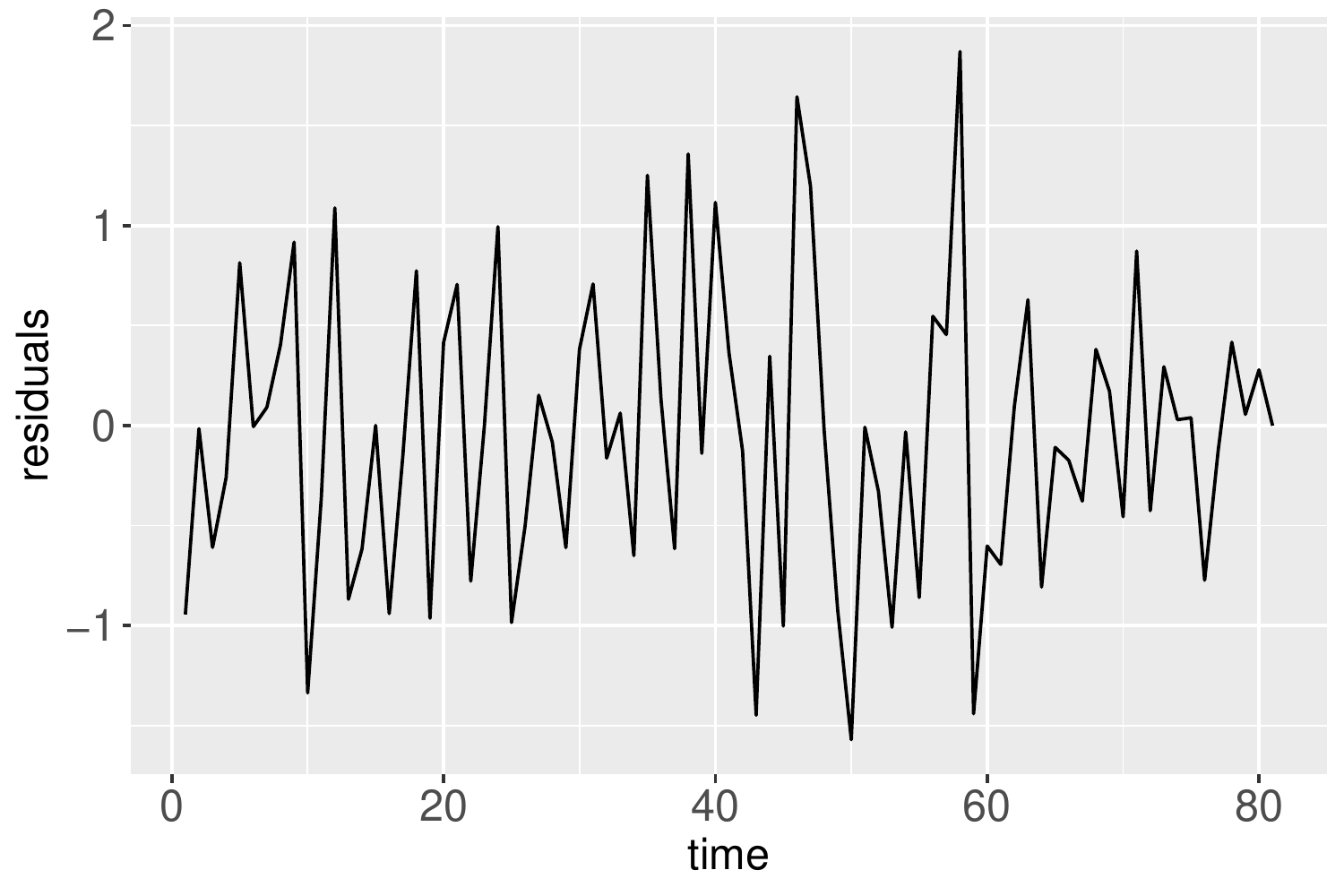}
    \caption{Residuals for plotted time series from fitting GNAR-edge model on \gr{the} sparsified network.}
        \label{res_hist_lines}   
\end{figure}

\subsection{Assessing model fit}
\begin{figure}[htb!]
    \centering
    \includegraphics[scale=.65]{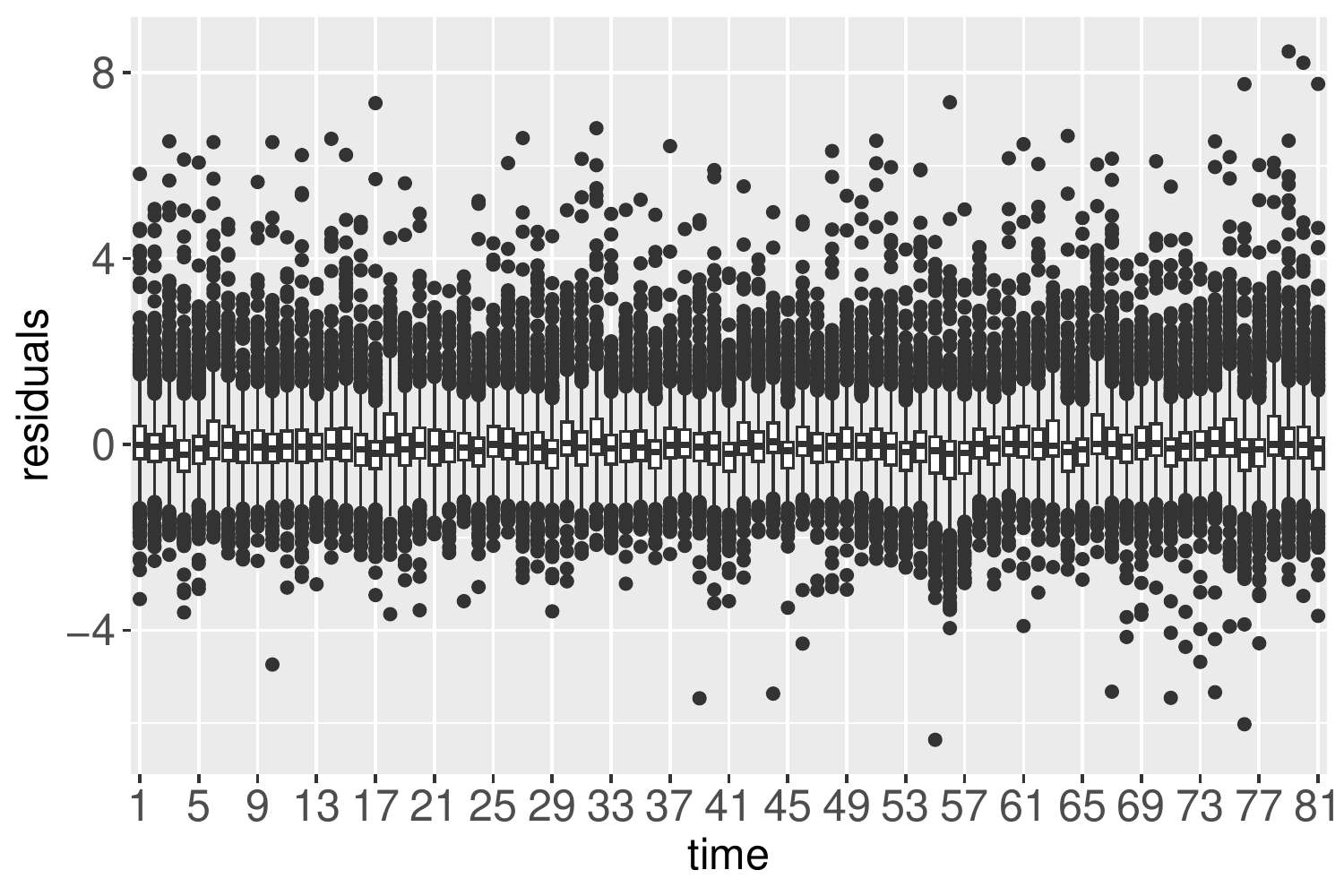}
    \caption{Distribution of residuals across time series for each time stamp ($x$ axis).}
        \label{res_box}   
\end{figure}

\begin{figure}[htb!]
    \centering
    \includegraphics[scale=.4]{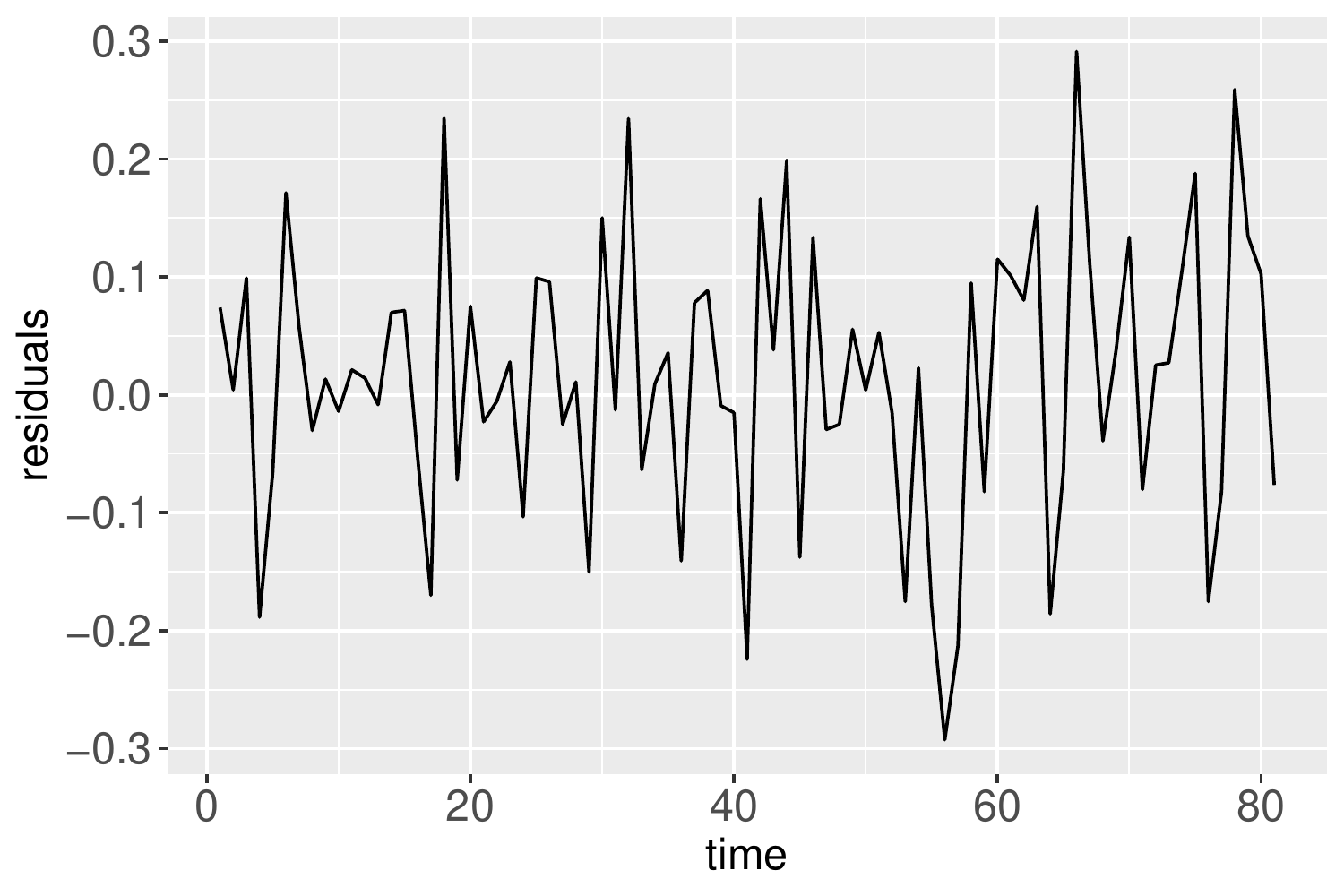}
    \qquad
    \centering
    \includegraphics[scale=.4]{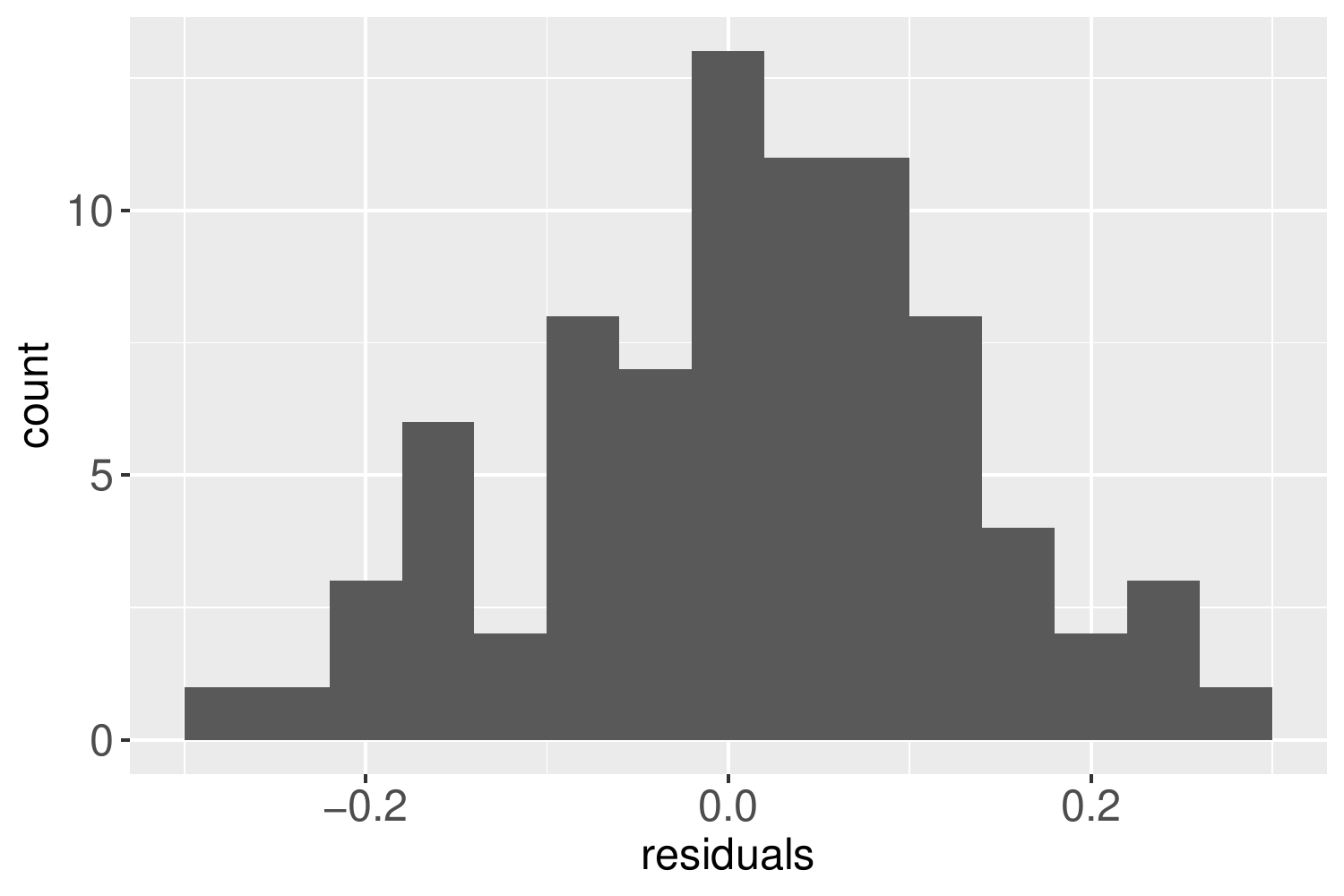}
\vfill
\centering
    \includegraphics[scale=.4]{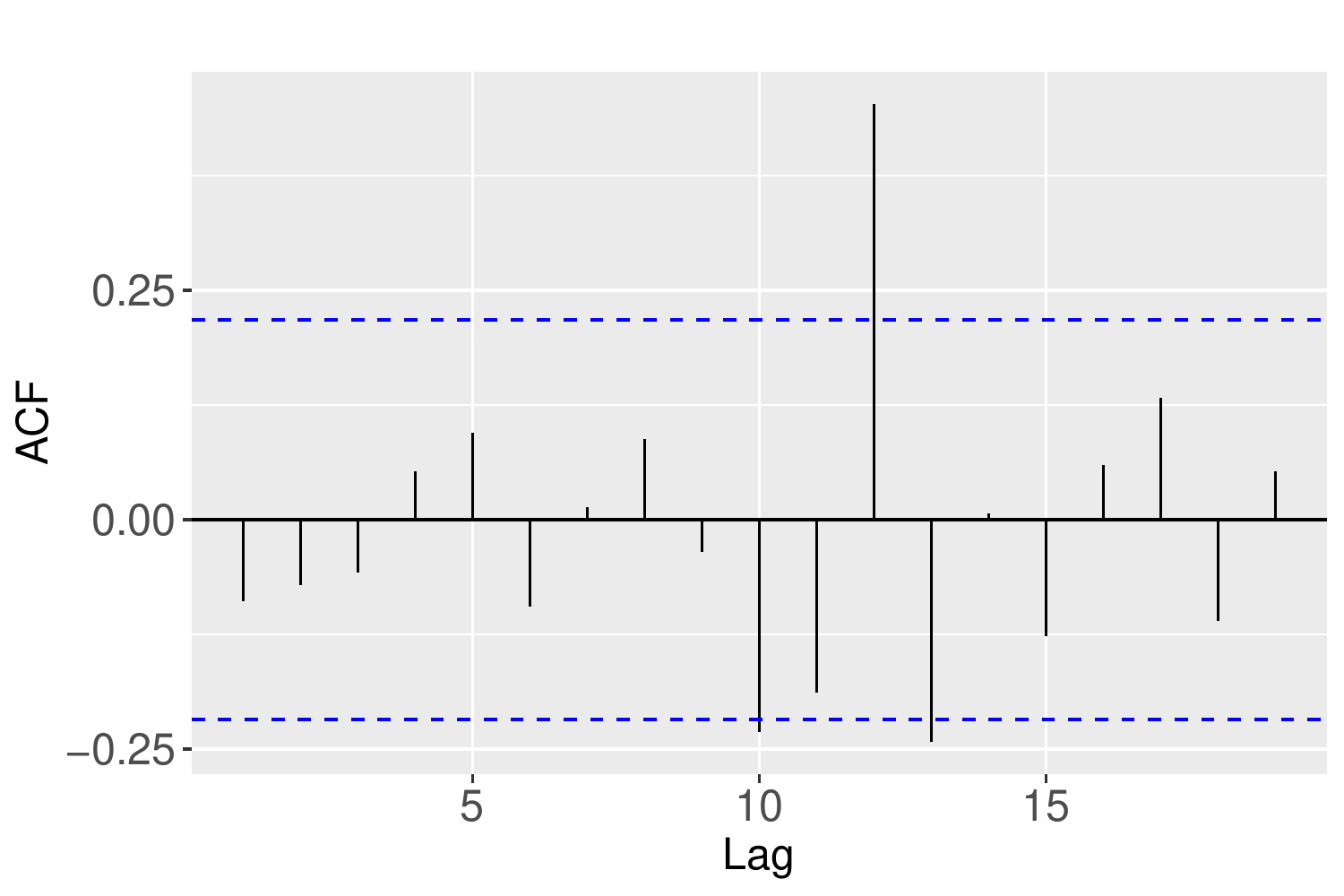}
    \qquad
    \centering
    \includegraphics[scale=.4]{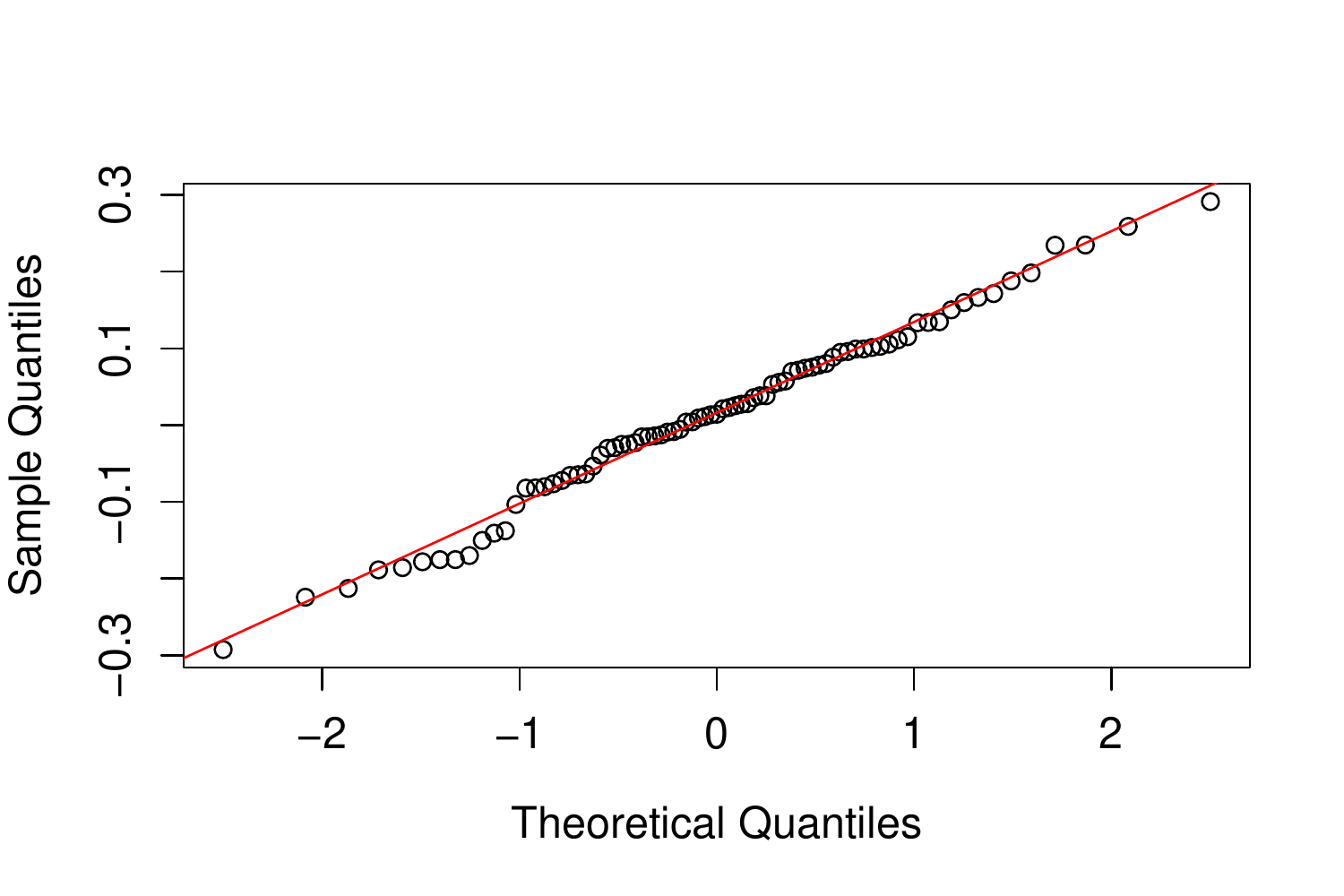}
    \caption{Mean residuals from fitting \gdr{the} GNAR-edge model on   \gdr{the} sparsified network.}
        \label{mean_res}   
\end{figure}

\gr{To assess model fit, we next} investigate the residuals from fitting the model on the sparsified network.
In Figure \ref{res_hist_lines}, we further plot the residuals for the edge time series of Figure \ref{fitt_lines}, respectively. 
In Figure \ref{res_hist_lines}, we further plot the residuals for the edge time series of Figure \ref{fitt_lines}, respectively. A challenge with performing \gdr{a full residual} 
analysis is that   
we have multiple  \gdr{(801)} time series. \gr{T}hus, we  \mcr{first  consider an approach} for summarising the residuals across the multiple time series. First, we plot the distribution of the residuals across the time series for each time stamp, shown in Figure \ref{res_box}. We observe that overall, residuals are distributed around 0, however, there are several outliers. We further consider the mean residual across the edge time series, for each time stamp, and plot the histogram, line, \gr{quantile-quantile} plot and autocorrelation of the resulting mean residuals as shown in Figure \ref{mean_res}. The plots suggest that the mean residuals 
\gdr{could be modelled as} normally distributed around 0, as also suggested by the Shapiro-Wilk normality test with $p$-value=0.78.

\begin{figure}[htb!]
    \centering
    \includegraphics[scale=.45]{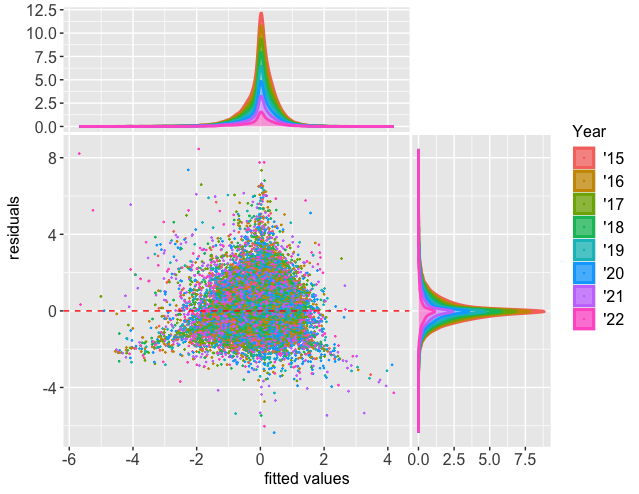}
    \caption{Fitted values (x axis) versus residuals (y axis) among multiple time series with points with same colour corresponding to year, for multiple time series.}
        \label{scat_den}   
\end{figure}

Lastly, we plot the fitted values versus the residuals in Figure \ref{scat_den}, with colours corresponding to different years. The side plots in Figure \ref{scat_den} \gdr{show}
the marginal distributions for the different years. 
\gdr{In view of the model assuming that the errors come from independent standard multivariate normal distributions},  \mcr{it is thus} encouraging to see that the marginal distribution exhibit a close to normal shape, as well as that there is not a specific pattern followed by the \gdr{residuals} 
across time.

\section{Conclusion}\label{sec7}
In this paper, we introduce a generalised network autoregressive model for networks with time varying edge weights,  \mcr{which we dub as}
\gdr{the GNAR-edge model}. \gdr{The GNAR-edge model} exploit\gdr{s} the underlying network structure of multivariate time series observed on the edges of the network. Synthetic data experiments validate the performance of \gdr{the GNAR-edge model} 
in parameter estimation and prediction. Comparisons to baseline models used for the analysis of multivariate time series systems show that 
\gdr{the GNAR-edge model can improve}
predictions when there is an underlying network structure in the data.

\gdr{The GNAR-edge model} is motivated by an anonymised and aggregated UK business payments data set made available to the Office for National Statistics by Pay.UK and Vocalink, which can be represented as a graph with time varying edge weights. The real network is a very densely connected network and thus we propose an approach for sparsifying the network using \gr{a} lead-lag analysis of the time series. Results show that \gdr{the GNAR-edge model} offers considerably more accurate predictions for the real data versus baseline models, especially after network sparsification.

An avenue for future work would be the investigation of communities formed by the edges of the network, and the inclusion of  
community structure information in the autoregressive model. In the network literature, there is vast research on the topic of identifying communities formed by the nodes of the network. On the other hand, the topic of communities \gr{formed}  
by the edges of the network has attracted relatively less attention 
(see \cite{suveges2022networks} and references therein). As in our example edges are associated with time series, \mcr{a viable theoretically-grounded  approach} to recover \mcr{the} underlying edge communities is to investigate the lead-lag relationships of the time series. Specifically, as introduced in \cite{bennett2022lead}, one can construct a lead-lag network from leading and lagging relationships between time series, and \mcr{subsequently} perform node clustering, which  
corresponds to time series clustering. One way to use the information from edge time series clustering is to incorporate it in the GNAR-edge model through weights \gdr{in \eqref{gen_gnar_edge}}, under the assumption that edges belonging to the same cluster  
have a greater effect on each other compared to edges belonging to different clusters.

Another interesting extension of our model would be the inclusion of node and edge covariates. Information on the nodes or the edges of a graph is commonly available in many real network applications. Incorporating  \mcr{such information in the model} in the form of covariates may \mcr{enhance the}  prediction performance.

\gdr{Furthermore, i}n 
\gdr{the GNAR-edge}  model, the network effect is encoded by a single parameter, for both incoming and outgoing neighbouring edges, for each edge. In certain applications, it might be of interest to encode the effect of incoming and outgoing edges separately, 
through two different parameters. This can be especially relevant to financial applications, and specifically payments data, where a payment from $i$ to $j$ may depend on payments received by $i$ and payments made by $i$ in a different way. This distinction could be introduced in \gdr{the GNAR-edge}  
and explored as future work.

\section*{Computing}

All experiments were run on MacBook Pro M1 machine.

\section*{Funding}

\gdr{The authors gratefully acknowledge  the ONS-Turing Strategic partnership for funding.}
Gesine Reinert was supported in part by the Engineering and Physical Sciences Research Council (EPSRC) grants [EP/T018445/1, EP/R018472/1, EP/X0021951].

\section*{Acknowledgment}

The authors 
\gdr{would like to thank the ONS-Turing Strategic Partnership team, in particular} Francois Lafond, Kerstin Hotte, \gdr{and} Johannes Lumma, 
for useful comments and economic insights to this project. We also thank Keith Lai, Alex Holmes, Dragos Cozma and Nathan Williams from \gdr{the Office for National Statistics}  
for providing an experimental version of the data, as well as for providing data insights. \gdr{Our gratitude also goes to 
Pay.UK and Vocalink for their support.} Finally we would like to thank an anonymous reviewer for their helpful comments and suggestions.

\appendix
\section{Appendix: Additional results for the synthetic data experiments}\label{App}

\subsection{Some network illustrations}\label{AppA1}

\begin{figure}[htb!]
    \centering
    \includegraphics[scale=.45]{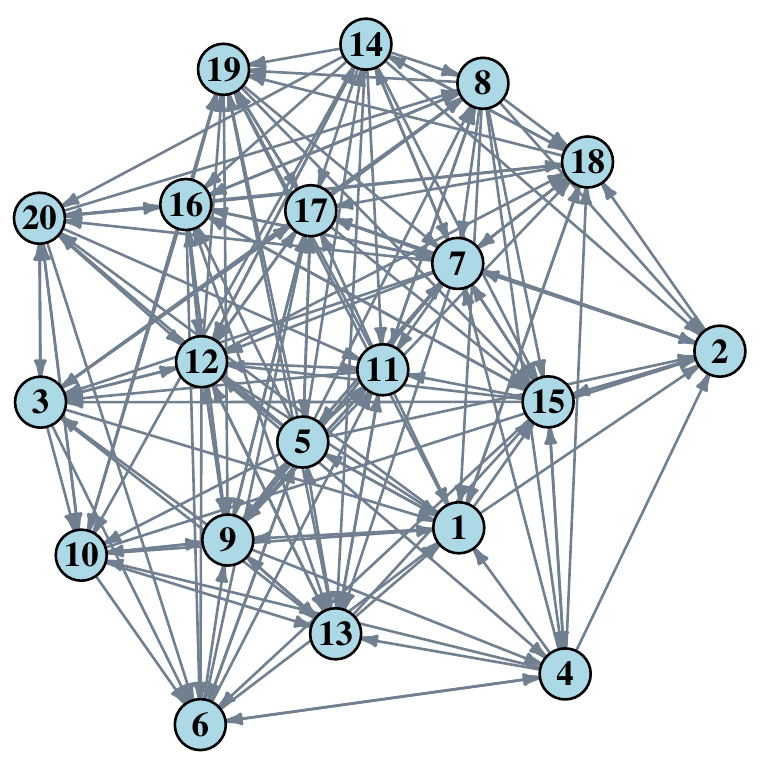}
    \hspace{-0.2cm}
        \centering
    \includegraphics[scale=.6]{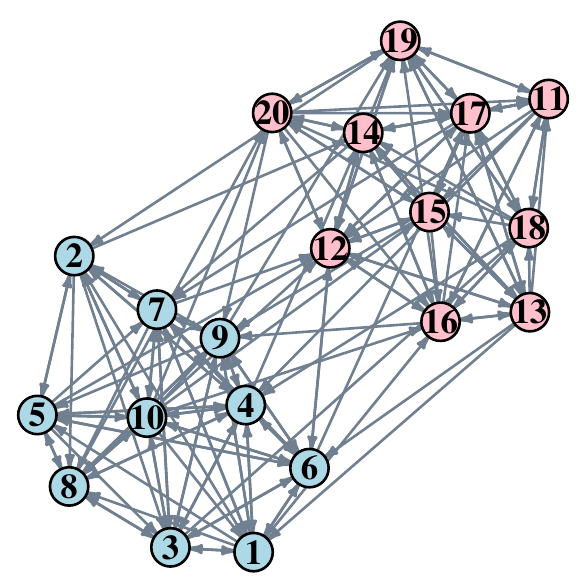}
    \hspace{-.2cm}
    \centering
    \includegraphics[scale=.7]{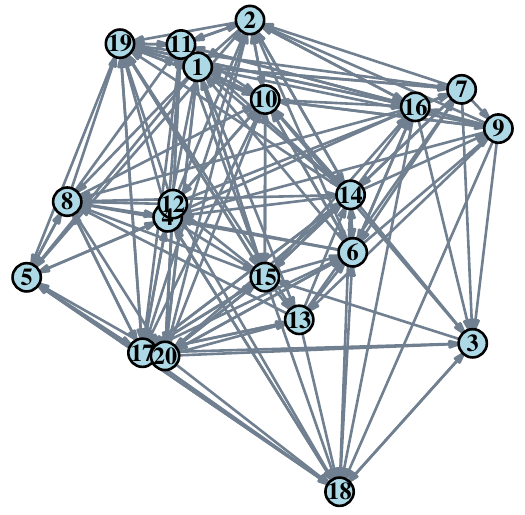}
    \caption{A \gr{simulated} ER network (left), SBM network (middle) and RDP network (right), \gr{for each of which time series on the edges are simulated}.}
    \label{er_sbm_rdp}
\end{figure}

In Figure \ref{er_sbm_rdp}, we present a network instance for each network structure considered (ER, SBM, RDP) for synthetic data experiments of Section \ref{sec511}.

\newpage
\subsection{The effect of network density on the 
predictive performance for moderately-sized networks}\label{AppA2}

\begin{figure}[htb!]
\centering
\begin{subfigure}[b]{.4\textwidth}
    \centering
    \includegraphics[scale=.35]{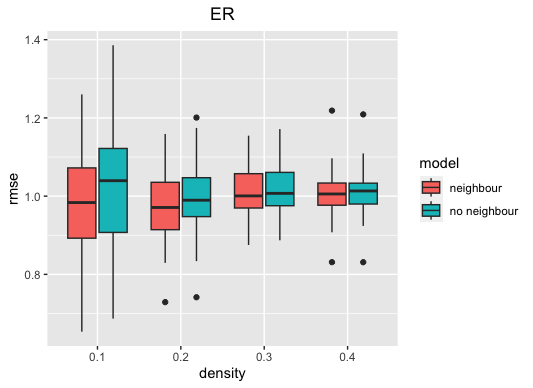}
    \caption{Distribution of RMSE for various network densities ($x$ axis) and GNAR-edge model with and without neighbour structure, for Erd\"{o}s-R\'{e}nyi networks.}
    \label{rmse_sim}
    \end{subfigure}
    \hspace{1cm}
    \begin{subfigure}[b]{.4\textwidth}
    \centering
    \includegraphics[scale=.35]{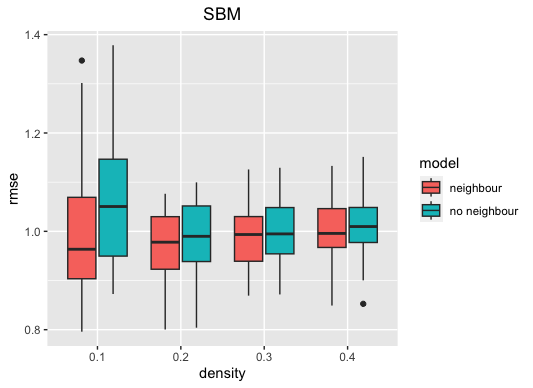}
    \caption{Distribution of RMSE for various network densities ($x$ axis) and GNAR-edge model with and without neighbour structure, for SBM networks.}
    \label{rmse_sim2}
    \end{subfigure}
    \vfill
    \begin{subfigure}[b]{.9\textwidth}
    \centering
    \includegraphics[scale=.35]{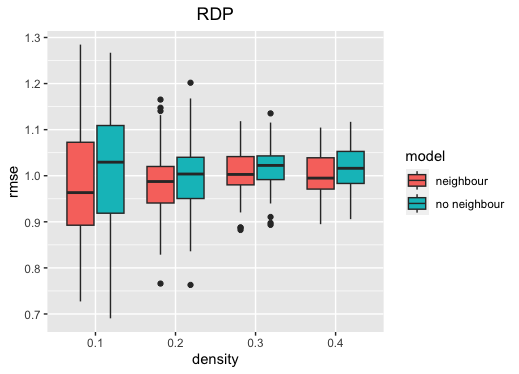}
    \caption{Distribution of RMSE for various network densities ($x$ axis) and GNAR-edge model with and without neighbour structure, for RDP networks.}
    \label{rmse_sim3}
    \end{subfigure}
    \caption{Distribution of RMSE for various network densities ($x$ axis) and network structures (subfigures), after fitting the GNAR-edge model with and without neighbour structure.}\label{dens}
\end{figure}

In Figure \ref{dens}  we report the results obtained from investigating the predictive performance of the GNAR-edge model with and without network structure, for different network densities, as discussed in Section \ref{sec521}.

\newpage

\subsection{Comparisons of predictive performance for large networks}\label{AppA3}
\begin{figure}[htb!]
    \centering
    \includegraphics[scale=.5]{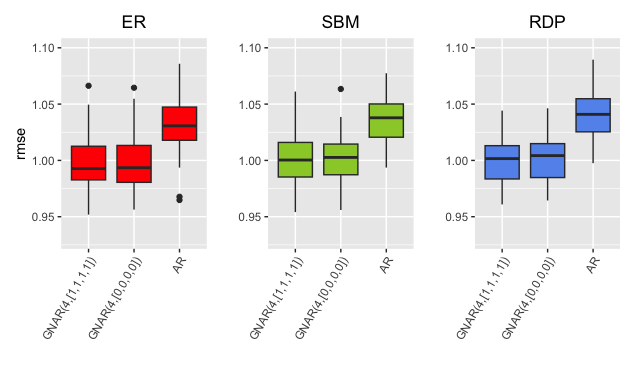}
    \caption{Distribution of RMSE for different structures Erd\"{o}s-R\'{e}nyi, SBM and RDP, for 86-node networks with density approximately 0.1.}
    \label{rmse_sim_large}
\end{figure}

\begin{figure}[htb!]
    \centering
    \includegraphics[scale=.43]{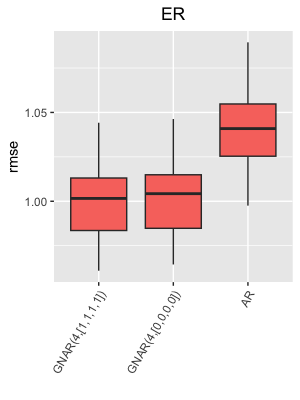}
    \caption{Distribution of RMSE for Erd\"{o}s-R\'{e}nyi network structure, for 86-node networks with density approximately 0.9.}
    \label{rmse_sim_large2}
\end{figure}

In this section, Figures \ref{rmse_sim_large} and \ref{rmse_sim_large2} show results from comparing the predictive performance of the GNAR-edge model and a simple AR model, for large networks, as discussed in Section \ref{sec522}.

\subsection{Additional results for model misspecification for moderately-sized networks}\label{AppA4}
In this section we present the results for the estimation and predictive performance of the GNAR-edge model under different model misspecification scenarios.

\begin{figure}[htb!]
    \centering
    \includegraphics[scale=.45]{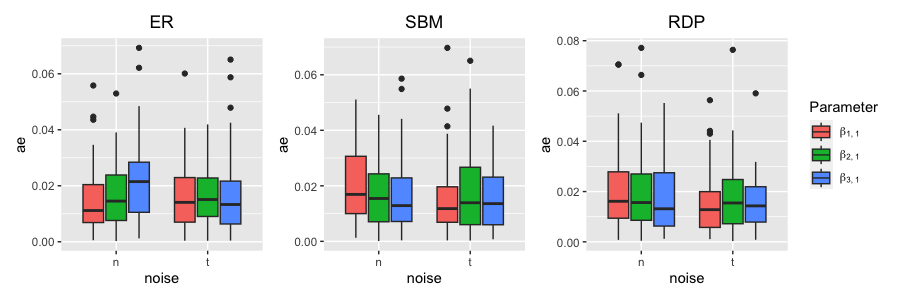}
    \caption{Distribution of absolute error for network effect parameters $\beta_{1,1},\beta_{2,1},\beta_{3,1}$, for different network structures, and normally distributed (n) and t-distributed with 3 degrees of freedom (t) innovations.}
    \label{est_ae_betas1}
\end{figure}

\begin{figure}[htb!]
    \centering
    \includegraphics[scale=.45]{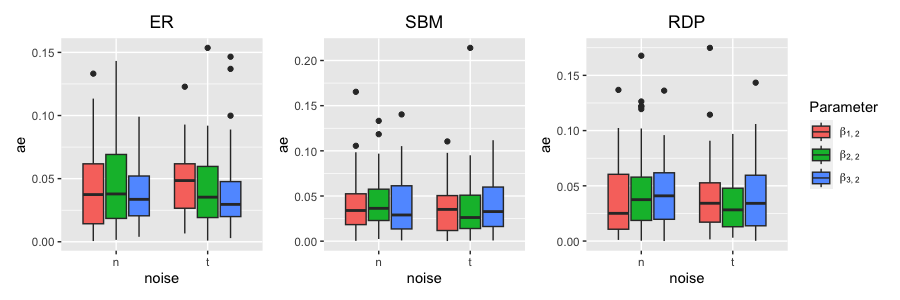}
    \caption{Distribution of absolute error for network effect parameters $\beta_{1,2},\beta_{2,2},\beta_{3,2}$, for different network structures, and normally distributed (n) and t-distributed with 3 degrees of freedom (t) innovations.}
    \label{est_ae_betas2}
\end{figure}

First, Figures \ref{est_ae_betas1} and \ref{est_ae_betas2} show the distribution of the absolute error of the estimated parameters for the network effect, considering normally distributed innovations (n) and t-distributed innovations with 3 degrees of freedom (t), as discussed in Section \ref{sec531}.

\begin{figure}[htb!]
    \centering
    \includegraphics[scale=.45]{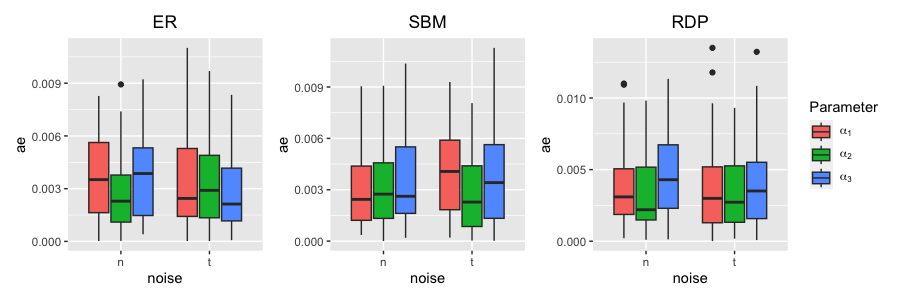}
    \caption{Distribution of absolute error for autoregressive parameters $\alpha_1,\alpha_2,\alpha_3$, for different network structures, and normally distributed (n) and t-distributed with 10 degrees of freedom (t) innovations.}
    \label{est_ae_alphast10}
\end{figure}

\begin{figure}[htb!]
    \centering
    \includegraphics[scale=.45]{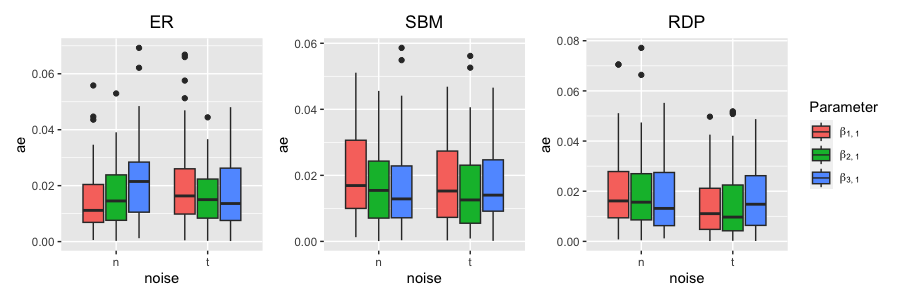}
    \caption{Distribution of absolute error for network effect parameters $\beta_{1,1},\beta_{2,1},\beta_{3,1}$, for different network structures, and normally distributed (n) and t-distributed with 10 degrees of freedom (t) innovations.}
    \label{est_ae_betas1_10}
\end{figure}

\begin{figure}[htb!]
    \centering
    \includegraphics[scale=.45]{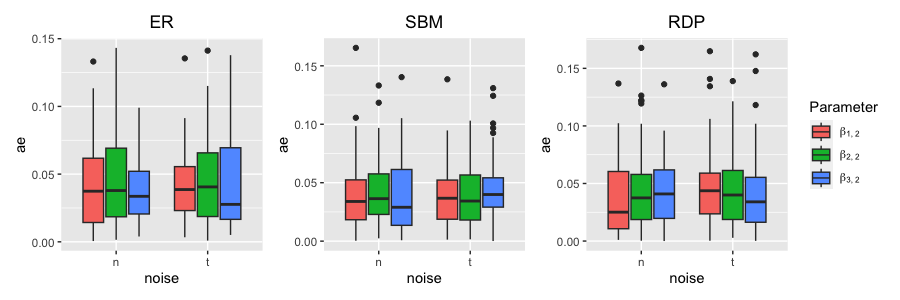}
    \caption{Distribution of absolute error for network effect parameters $\beta_{1,2},\beta_{2,2},\beta_{3,2}$, for different network structures, and normally distributed (n) and t-distributed with 10 degrees of freedom (t) innovations.}
    \label{est_ae_betas2_10}
\end{figure}

\begin{figure}[htb!]
    \centering
    \includegraphics[scale=.5]{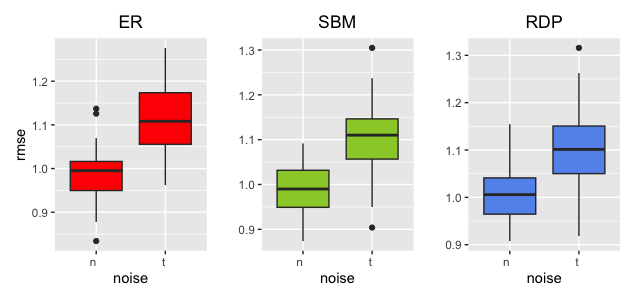}
    \caption{Distribution of RMSE for normally distributed (n) and t-distributed with 10 degrees of freedom (t) innovations, for different network structures.}
    \label{rmse_noiset10}
\end{figure}

Similarly Figures \ref{est_ae_alphast10}, \ref{est_ae_betas1_10}, \ref{est_ae_betas2_10} and \ref{rmse_noiset10} show the estimation and prediction performance for network time series simulated with normally distributed and t-distributed innovations with 10 degrees of freedom.


\clearpage

\begin{figure}[htb!]
    \centering
    \includegraphics[scale=.42]{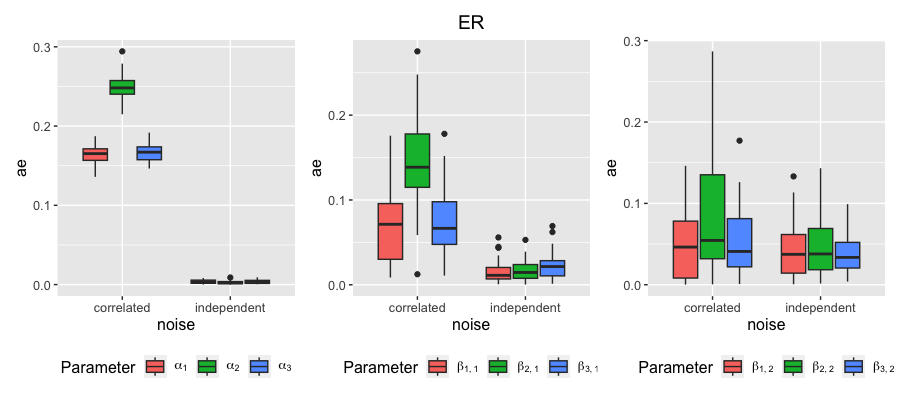}
    \caption{Distribution of absolute error for all model parameters for settings with independent and correlated innovations with correlation 0.5, for ER network structure.}
    \label{est_ae_allpar_corr_er}
\end{figure}

\begin{figure}[htb!]
    \centering
    \includegraphics[scale=.4]{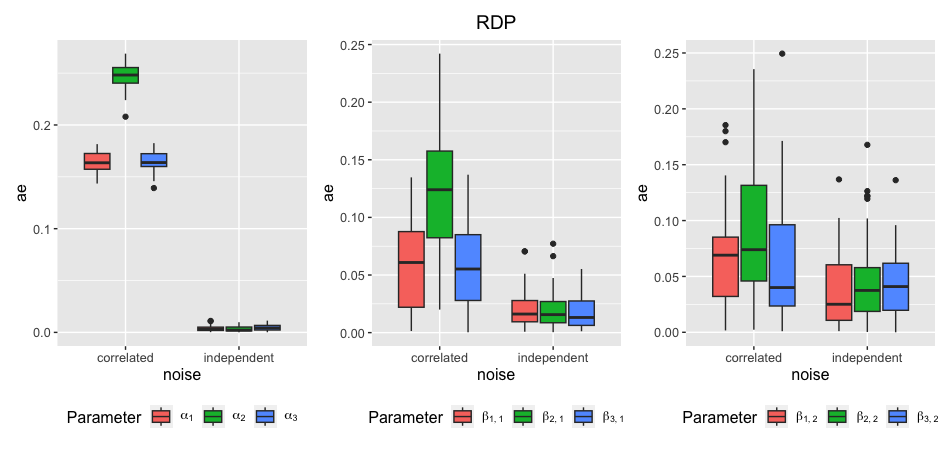}
    \caption{Distribution of absolute error for all model parameters for settings with independent and correlated innovations with correlation 0.5, for RDP network structure.}
    \label{est_ae_allpar_corr_rdp}
\end{figure}

In Figures \ref{est_ae_allpar_corr_er} and \ref{est_ae_allpar_corr_rdp}, we present the results for the estimation performance of the GNAR-edge model for i.i.d. innovations generated from multivariate standard normal distribution and correlated innovations generated from multivariate standard normal distribution with correlation 0.5, for ER and RDP networks, as discussed in Section \ref{sec531}.

\begin{figure}[htb!]
    \centering
    \includegraphics[scale=.42]{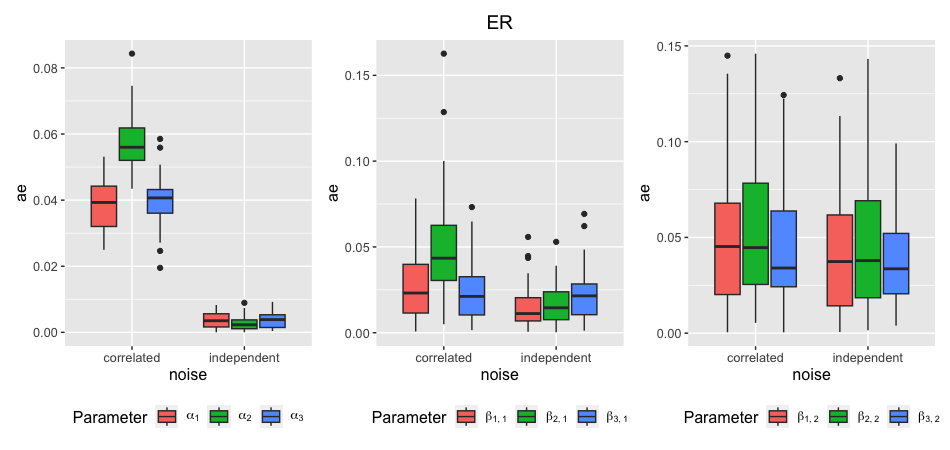}
    \caption{Distribution of absolute error for all model parameters for settings with independent and correlated innovations with correlation 0.1, for ER network structure.}
    \label{est_ae_allpar_corr0.1_er}
\end{figure}

\begin{figure}[htb!]
    \centering
    \includegraphics[scale=.4]{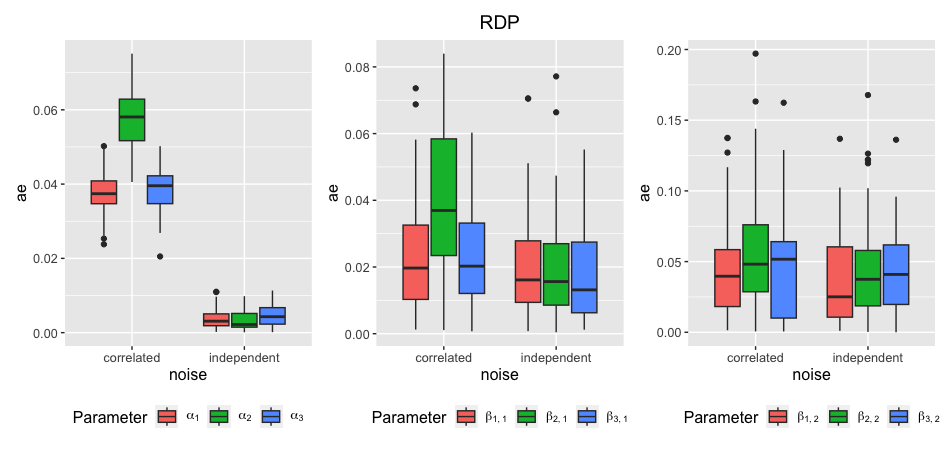}
    \caption{Distribution of absolute error for all model parameters for settings with independent and correlated innovations with correlation 0.1, for RDP network structure.}
    \label{est_ae_allpar_corr0.1_rdp}
\end{figure}

\begin{figure}[htb!]
    \centering
    \includegraphics[scale=.4]{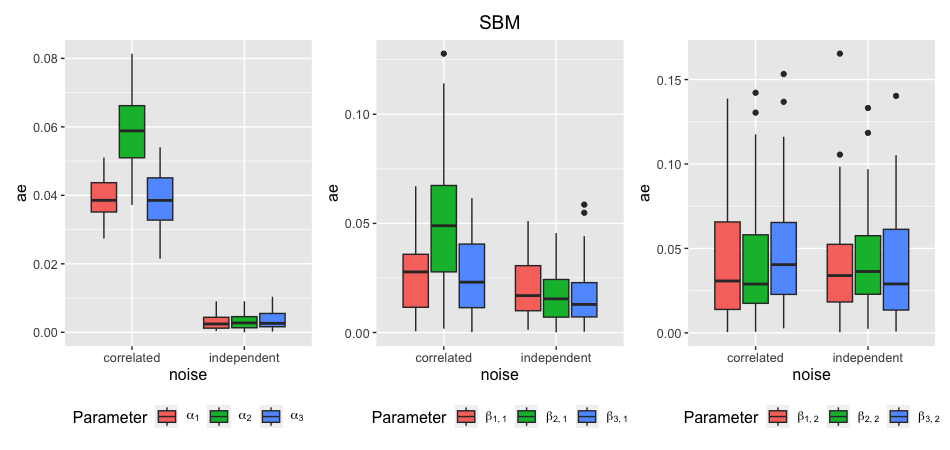}
    \caption{Distribution of absolute error for all model parameters for settings with independent and correlated innovations with correlation 0.1, for SBM network structure.}
    \label{est_ae_allpar_corr0.1_sbm}
\end{figure}

\newpage
We further perform the same experiments for independent and correlated innovations with correlation 0.1, as discussed in Section \ref{sec531}. The estimation performance of the GNAR-edge model is presented in Figures \ref{est_ae_allpar_corr0.1_er}, \ref{est_ae_allpar_corr0.1_rdp} and \ref{est_ae_allpar_corr0.1_sbm}.

\newpage
\subsection{Additional results for model misspecification for large networks}\label{AppA5}

In this section we present results for synthetic data experiments performed under model misspecification settings for large networks, as discussed in Section \ref{sec532}.

\begin{figure}[htb!]
    \centering
    \includegraphics[scale=.45]{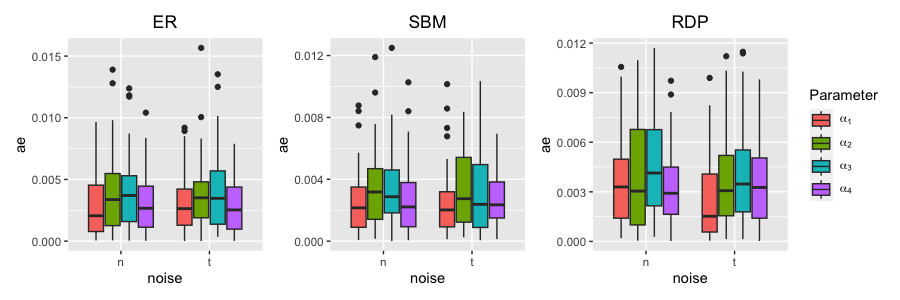}
    \caption{Distribution of absolute error for autoregressive parameters $\alpha_1,\alpha_2,\alpha_3,\alpha_4$, for different network structures, and normally distributed (n) and t-distributed with 3 degrees of freedom (t) innovations.}
    \label{est_ae_alphas_large}
\end{figure}

\begin{figure}[htb!]
    \centering
    \includegraphics[scale=.45]{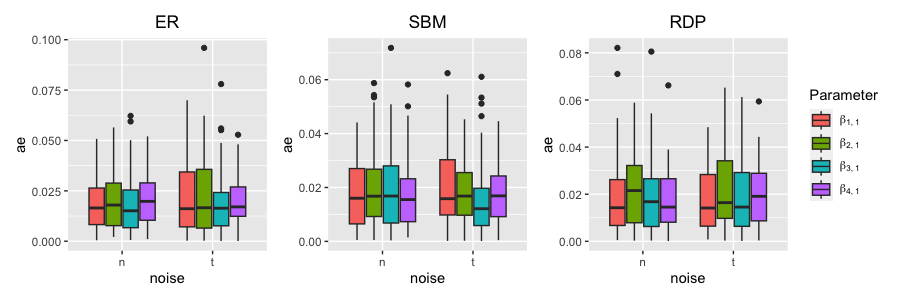}
    \caption{Distribution of absolute error for network effect parameters $\beta_{1,1},\beta_{2,1},\beta_{3,1},\beta_{4,1}$, for different network structures, and normally distributed (n) and t-distributed with 3 degrees of freedom (t) innovations.}
    \label{est_ae_betas1_large}
\end{figure}

\begin{figure}[htb!]
    \centering
    \includegraphics[scale=.5]{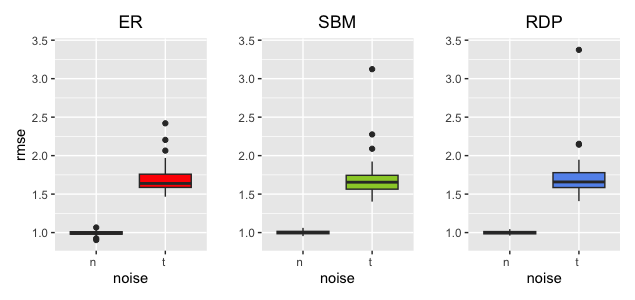}
    \caption{Distribution of RMSE for normally distributed (n) and t-distributed with 3 degrees of freedom (t) innovations, for different network structures.}
    \label{rmse_noise_large}
\end{figure}

Figures \ref{est_ae_alphas_large} and \ref{est_ae_betas1_large} show the distribution of the absolute error of the estimated parameters, for synthetic data generated with normally distributed and t-distributed innovations with 3 degrees of freedom, for all network structures considered. Figure \ref{rmse_noise_large} shows the distribution of the RMSE for predicting the last time stamp, under normally distributed and t-distributed innovations with 3 degrees of freedom.

\clearpage
\newpage

\begin{figure}[htb!]
    \centering
    \includegraphics[scale=.4]{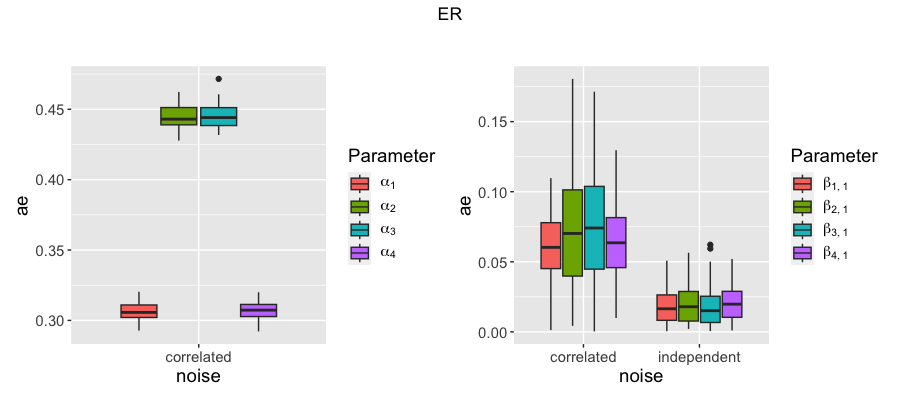}
    \caption{Distribution of absolute error for all model parameters for settings with independent and correlated innovations with correlation 0.5, for ER network structure.}
    \label{est_ae_all_er_corr_large}
\end{figure}

\begin{figure}[htb!]
    \centering
    \includegraphics[scale=.4]{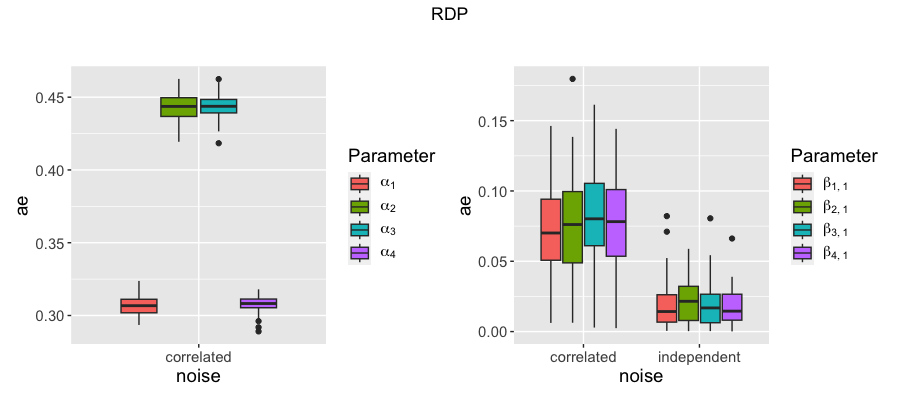}
    \caption{Distribution of absolute error for all model parameters for settings with independent and correlated innovations with correlation 0.5, for RDP network structure.}
    \label{est_ae_all_rdp_corr_large}
\end{figure}

Figures \ref{est_ae_all_er_corr_large} and \ref{est_ae_all_rdp_corr_large} show the estimation performance of the GNAR-edge model for independent and correlated innovations with correlation  0.5, for ER and RDP network structures.
Figures \ref{est_ae_all_er_corr_large} and \ref{est_ae_all_rdp_corr_large}, left panels, show only the distribution of the absolute error for the correlated errors setting as colours are indistinguishable when independent errors setting is included in same plot, due to difference in the scale of the absolute error. For the results for the autoregressive parameters under the independent errors setting, see Figures \ref{est_ae_alphas_large}, \ref{est_ae_betas1_large} (n) case.

\begin{figure}[htb!]
    \centering
    \includegraphics[scale=.47]{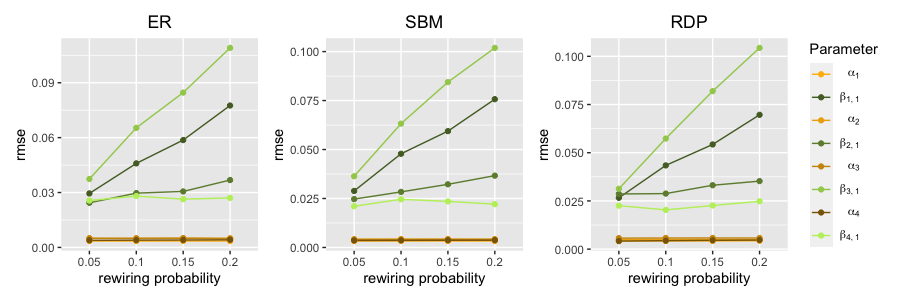}
    \caption{RMSE of estimated model parameters after 50 repetitions, and various perturbations of network structure (x axis).}
    \label{rmse_est_rewiring_large}
\end{figure}

\clearpage
\newpage
Figure \ref{rmse_est_rewiring_large} shows the RMSE for all mdoel parameters and network structures under misspecification of the network structure through edge rewiring, for different rewiring probabilities.

\bibliography{sample}

\begin{thebibliography}{}

\bibitem[Abdi and Amrit, 2021]{abdi2021review}
Abdi, A. and Amrit, C. (2021).
\newblock A review of travel and arrival-time prediction methods on road
  networks: classification, challenges and opportunities.
\newblock {\em PeerJ Computer Science}, 7:e689.

\bibitem[Armillotta and Fokianos, 2021]{armillotta2021poisson}
Armillotta, M. and Fokianos, K. (2021).
\newblock Poisson network autoregression.
\newblock {\em arXiv preprint arXiv:2104.06296}.

\bibitem[Bennett et~al., 2022]{bennett2022lead}
Bennett, S., Cucuringu, M., and Reinert, G. (2022).
\newblock Lead--lag detection and network clustering for multivariate time
  series with an application to the us equity market.
\newblock {\em Machine Learning}, pages 1--42.

\bibitem[Bernanke et~al., 2005]{bernanke2005measuring}
Bernanke, B.~S., Boivin, J., and Eliasz, P. (2005).
\newblock Measuring the effects of monetary policy: a factor-augmented vector
  autoregressive (favar) approach.
\newblock {\em The Quarterly Journal of Economics}, 120(1):387--422.

\bibitem[Box et~al., 2015]{box2015time}
Box, G.~E., Jenkins, G.~M., Reinsel, G.~C., and Ljung, G.~M. (2015).
\newblock {\em Time Series Analysis: Forecasting and Control}.
\newblock John Wiley \& Sons.

\bibitem[Buda et~al., 2022]{buda2022national}
Buda, G., Hansen, S., Rodrigo, T., Carvalho, V., Ortiz, {\'A}., and Mora, J.
  V.~R. (2022).
\newblock National accounts in a world of naturally occurring data: A proof of
  concept for consumption.
\newblock {\em Cambridge Working Papers, Faculty of Economics, University of
  Cambridge}.

\bibitem[Carvalho et~al., 2021]{carvalho2021tracking}
Carvalho, V.~M., Garcia, J.~R., Hansen, S., Ortiz, {\'A}., Rodrigo, T.,
  Rodr{\'\i}guez~Mora, J.~V., and Ruiz, P. (2021).
\newblock Tracking the covid-19 crisis with high-resolution transaction data.
\newblock {\em Royal Society Open Science}, 8(8):210218.

\bibitem[Chen et~al., 2023]{chen2022community}
Chen, E.~Y., Fan, J., and Zhu, X. (2023).
\newblock Community network auto-regression for high-dimensional time series.
\newblock {\em Journal of Econometrics}, 235(2):1239--1256.

\bibitem[{Companies House}, 2018]{sic}
{Companies House}, U.~K. (2018).
\newblock Standard industrial classification of economic activities ({SIC}).
\newblock
  \url{https://www.gov.uk/government/publications/standard-industrial-classification-of-economic-activities-sic}.

\bibitem[Csardi and Nepusz, 2006]{igraph}
Csardi, G. and Nepusz, T. (2006).
\newblock The igraph software package for complex network research.
\newblock {\em InterJournal}, Complex Systems:1695.

\bibitem[Dallakyan et~al., 2022]{dallakyan2022time}
Dallakyan, A., Kim, R., and Pourahmadi, M. (2022).
\newblock Time series graphical lasso and sparse var estimation.
\newblock {\em Computational Statistics \& Data Analysis}, 176:107557.

\bibitem[Davis et~al., 2016]{davis2016sparse}
Davis, R.~A., Zang, P., and Zheng, T. (2016).
\newblock Sparse vector autoregressive modeling.
\newblock {\em Journal of Computational and Graphical Statistics},
  25(4):1077--1096.

\bibitem[de~la Torre et~al., 2016]{de2016topologic}
de~la Torre, S.~R., Kalda, J., Kitt, R., and Engelbrecht, J. (2016).
\newblock On the topologic structure of economic complex networks: empirical
  evidence from large scale payment network of {E}stonia.
\newblock {\em Chaos, Solitons \& Fractals}, 90:18--27.

\bibitem[Durante and Dunson, 2016]{durante2016locally}
Durante, D. and Dunson, D.~B. (2016).
\newblock Locally adaptive dynamic networks.
\newblock {\em The Annals of Applied Statistics}, 10(4):2203--2232.

\bibitem[Fu et~al., 2009]{fu2009dynamic}
Fu, W., Song, L., and Xing, E.~P. (2009).
\newblock Dynamic mixed membership blockmodel for evolving networks.
\newblock In {\em Proceedings of the 26th annual International Conference on
  Machine Learning}, pages 329--336.

\bibitem[Hanneke et~al., 2010]{hanneke2010discrete}
Hanneke, S., Fu, W., and Xing, E.~P. (2010).
\newblock Discrete temporal models of social networks.
\newblock {\em Electronic Journal of Statistics}, 4:585--605.

\bibitem[Jiang et~al., 2020]{jiang2020autoregressive}
Jiang, B., Li, J., and Yao, Q. (2020).
\newblock Autoregressive networks.
\newblock {\em arXiv preprint arXiv:2010.04492}.

\bibitem[Kang et~al., 2021]{kang2021dynamic}
Kang, X., Ganguly, A., and Kolaczyk, E.~D. (2021).
\newblock Dynamic networks with multi-scale temporal structure.
\newblock {\em Sankhya A}, pages 1--43.

\bibitem[Knight et~al., 2020]{knight2019generalised}
Knight, M., Leeming, K., Nason, G., and Nunes, M. (2020).
\newblock Generalised network autoregressive processes and the {GNAR} package.
\newblock {\em Journal of Statistical Software}, 96(5):1--36.

\bibitem[Koop, 2013]{koop2013forecasting}
Koop, G.~M. (2013).
\newblock Forecasting with medium and large bayesian vars.
\newblock {\em Journal of Applied Econometrics}, 28(2):177--203.

\bibitem[Krivitsky and Handcock, 2014]{krivitsky2014separable}
Krivitsky, P.~N. and Handcock, M.~S. (2014).
\newblock A separable model for dynamic networks.
\newblock {\em Journal of the Royal Statistical Society. Series B, Statistical
  Methodology}, 76(1):29.

\bibitem[Ludkin et~al., 2018]{ludkin2018dynamic}
Ludkin, M., Eckley, I., and Neal, P. (2018).
\newblock Dynamic stochastic block models: parameter estimation and detection
  of changes in community structure.
\newblock {\em Statistics and Computing}, 28:1201--1213.

\bibitem[L{\"u}tkepohl, 2005]{lutkepohl2005new}
L{\"u}tkepohl, H. (2005).
\newblock {\em New introduction to multiple time series analysis}.
\newblock Springer Science \& Business Media.

\bibitem[Matias and Miele, 2017]{matias2017statistical}
Matias, C. and Miele, V. (2017).
\newblock Statistical clustering of temporal networks through a dynamic
  stochastic block model.
\newblock {\em Journal of the Royal Statistical Society. Series B (Statistical
  Methodology)}, 79(4):1119--1141.

\bibitem[Menelaou et~al., 2018]{menelaou2018effective}
Menelaou, C., Kolios, P., Timotheou, S., and Panayiotou, C.~G. (2018).
\newblock Effective prediction of road segment occupancy for the
  route-reservation architecture.
\newblock {\em IFAC-PapersOnLine}, 51(9):470--475.

\bibitem[Min and Wynter, 2011]{min2011real}
Min, W. and Wynter, L. (2011).
\newblock Real-time road traffic prediction with spatio-temporal correlations.
\newblock {\em Transportation Research Part C: Emerging Technologies},
  19(4):606--616.

\bibitem[Nicholson et~al., 2020]{nicholson2020high}
Nicholson, W.~B., Wilms, I., Bien, J., and Matteson, D.~S. (2020).
\newblock High dimensional forecasting via interpretable vector autoregression.
\newblock {\em The Journal of Machine Learning Research}, 21(1):6690--6741.

\bibitem[Pensky, 2019]{pensky2019dynamic}
Pensky, M. (2019).
\newblock Dynamic network models and graphon estimation.
\newblock {\em The Annals of Statistics}, 47(4):2378--2403.

\bibitem[Salamanis et~al., 2015]{salamanis2015managing}
Salamanis, A., Kehagias, D.~D., Filelis-Papadopoulos, C.~K., Tzovaras, D., and
  Gravvanis, G.~A. (2015).
\newblock Managing spatial graph dependencies in large volumes of traffic data
  for travel-time prediction.
\newblock {\em IEEE Transactions on Intelligent Transportation Systems},
  17(6):1678--1687.

\bibitem[Sarkar and Moore, 2005]{sarkar2005dynamic}
Sarkar, P. and Moore, A.~W. (2005).
\newblock Dynamic social network analysis using latent space models.
\newblock {\em ACM SIGKDD Explorations Newsletter}, 7(2):31--40.

\bibitem[Sioofy~Khoojine et~al., 2021]{sioofy2021network}
Sioofy~Khoojine, A., Shadabfar, M., Hosseini, V.~R., and Kordestani, H. (2021).
\newblock Network autoregressive model for the prediction of covid-19
  considering the disease interaction in neighboring countries.
\newblock {\em Entropy}, 23(10):1267.

\bibitem[Spencer et~al., 2015]{spencer2015inferring}
Spencer, S.~E., Hill, S.~M., and Mukherjee, S. (2015).
\newblock Inferring network structure from interventional time-course
  experiments.
\newblock {\em The Annals of Applied Statistics}, 9(1):507--524.

\bibitem[Suveges and Olhede, 2023]{suveges2022networks}
Suveges, M. and Olhede, S.~C. (2023).
\newblock Networks with correlated edge processes.
\newblock {\em Journal of the Royal Statistical Society Series A: Statistics in
  Society}, 186:441--462.

\bibitem[Tsay, 2013]{tsay2013multivariate}
Tsay, R.~S. (2013).
\newblock {\em Multivariate time series analysis: with R and financial
  applications}.
\newblock John Wiley \& Sons.

\bibitem[Wu et~al., 2010]{wu2010detecting}
Wu, D., Ke, Y., Yu, J.~X., Yu, P.~S., and Chen, L. (2010).
\newblock Detecting leaders from correlated time series.
\newblock In {\em Database Systems for Advanced Applications: 15th
  International Conference, DASFAA 2010, Tsukuba, Japan, April 1-4, 2010,
  Proceedings, Part I 15}, pages 352--367. Springer.

\bibitem[Zhu et~al., 2020]{zhu2020multivariate}
Zhu, X., Huang, D., Pan, R., and Wang, H. (2020).
\newblock Multivariate spatial autoregressive model for large scale social
  networks.
\newblock {\em Journal of Econometrics}, 215(2):591--606.

\bibitem[Zhu et~al., 2017]{zhu2017network}
Zhu, X., Pan, R., Li, G., Liu, Y., and Wang, H. (2017).
\newblock Network vector autoregression.
\newblock {\em The Annals of Statistics}, 45(3):1096--1123.

\bibitem[Zhu et~al., 2019]{zhu2019network}
Zhu, X., Wang, W., Wang, H., and H{\"a}rdle, W.~K. (2019).
\newblock Network quantile autoregression.
\newblock {\em Journal of Econometrics}, 212(1):345--358.

\end{thebibliography}

\end{document}


\title{Supplementary Material: \gr{The} GNAR-edge model: A network autoregressive model for networks with time-varying edge weights}

\author[1]{Anastasia Mantziou}
\author[1,2]{Mihai Cucuringu}
\author[3]{Victor Meirinhos}
\author[1,2]{Gesine Reinert}
\affil[1]{The Alan Turing Institute}
\affil[2]{University of Oxford, Department of Statistics}
\affil[3]{Office for National Statistics}

\maketitle

In this document we provide supplementary material to the article "\gr{The} GNAR-edge model: A network autoregressive model for networks with time-varying edge weights". In Section 1, we provide additional results from synthetic data experiments presented in Section 5 of the main article. In Section 2, we report additional results for the real data applications presented in Section 6 of the main article.

\section{Additional details for synthetic data experiments}
In this section we provide additional results for the synthetic data experiments presented in Section 5.1.1 of the main article. Specifically, Tables \ref{covsim1}-\ref{covsim5} show the coverage rates for 95\% confidence intervals enclosing the true value of the parameters, as well as the Root Mean Square Error (RMSE) for the estimated coefficients with respect to their true values, for \gr{the} simulation regimes 1, 2, 3 and 5 of Table 1 in the main article.

\begin{table}[htb!]
\centering
\begin{tabular}{ccccccc}
         & \multicolumn{6}{c}{Regime 1}                                               \\ \cline{2-7} 
         & \multicolumn{2}{c}{ER} & \multicolumn{2}{c}{SBM} & \multicolumn{2}{c}{RDP} \\ \cline{2-7} 
         & \begin{tabular}[c]{@{}c@{}}Coverage\\ (\%)\end{tabular}    & RMSE     & \begin{tabular}[c]{@{}c@{}}Coverage\\ (\%)\end{tabular}    & RMSE     & \begin{tabular}[c]{@{}c@{}}Coverage\\ (\%)\end{tabular}     & RMSE     \\ \hline
$\alpha_1$ & 100         & 0.004    & 0.96         & 0.005    & 0.94         & 0.005    \\
$\beta_{1,1}$ & 0.94        & 0.03     & 0.9          & 0.03     & 0.92         & 0.03     \\ \hline
\end{tabular}
\caption{Coverage rates for $95\%$ confidence interval enclosing the true value and RMSE for estimated coefficients, across 50 replications for Simulation Regime 1.}
\label{covsim1}
\end{table}

\begin{table}[htb!]
\centering
\begin{tabular}{ccccccc}
         & \multicolumn{6}{c}{Regime 2}                                               \\ \cline{2-7} 
         & \multicolumn{2}{c}{ER} & \multicolumn{2}{c}{SBM} & \multicolumn{2}{c}{RDP} \\ \cline{2-7} 
         & \begin{tabular}[c]{@{}c@{}}Coverage\\ (\%)\end{tabular}    & RMSE     & \begin{tabular}[c]{@{}c@{}}Coverage\\ (\%)\end{tabular}     & RMSE     & \begin{tabular}[c]{@{}c@{}}Coverage\\ (\%)\end{tabular}     & RMSE     \\ \hline
$\alpha_1$ & 100         & 0.004    & 0.96         & 0.005    & 0.94         & 0.005    \\
$\beta_{1,1}$ & 0.94        & 0.03     & 0.88         & 0.03     & 0.9          & 0.03     \\
$\beta_{1,2}$ & 0.94        & 0.04     & 0.96         & 0.04     & 0.92         & 0.03     \\ \hline
\end{tabular}
\caption{Coverage rates for $95\%$ confidence interval enclosing the true value and RMSE for estimated coefficients, across 50 replications for Simulation Regime 2.}
\label{covsim2}
\end{table}

\begin{table}[htb!]
\centering
\begin{tabular}{ccccccc}
         & \multicolumn{6}{c}{Regime 3}                                               \\ \cline{2-7} 
         & \multicolumn{2}{c}{ER} & \multicolumn{2}{c}{SBM} & \multicolumn{2}{c}{RDP} \\ \cline{2-7} 
         & \begin{tabular}[c]{@{}c@{}}Coverage\\ (\%)\end{tabular}    & RMSE     & \begin{tabular}[c]{@{}c@{}}Coverage\\ (\%)\end{tabular}     & RMSE     & \begin{tabular}[c]{@{}c@{}}Coverage\\ (\%)\end{tabular}     & RMSE     \\ \hline
$\alpha_1$ & 100         & 0.004    & 0.98         & 0.003    & 0.94         & 0.004    \\
$\beta_{1,1}$ & 0.96        & 0.02     & 0.94         & 0.02     & 0.94         & 0.02     \\
$\alpha_2$ & 0.98        & 0.003    & 0.94         & 0.004    & 0.98         & 0.004    \\
$\beta_{2,1}$ & 0.96        & 0.02     & 0.98         & 0.02     & 0.92         & 0.02     \\
$\alpha_3$ & 0.98        & 0.004    & 0.96         & 0.004    & 0.9          & 0.005    \\
$\beta_{3,1}$ & 0.96        & 0.02     & 0.98         & 0.02     & 0.92         & 0.02     \\ \hline
\end{tabular}
\caption{Coverage rates for $95\%$ confidence interval enclosing the true value and RMSE for estimated coefficients, across 50 replications for Simulation Regime 3.}
\label{covsim3}
\end{table}

\begin{table}[htb!]
\centering
\begin{tabular}{ccccccc}
         & \multicolumn{6}{c}{Regime 5}                                               \\ \cline{2-7} 
         & \multicolumn{2}{c}{ER} & \multicolumn{2}{c}{SBM} & \multicolumn{2}{c}{RDP} \\ \cline{2-7} 
         & \begin{tabular}[c]{@{}c@{}}Coverage\\ (\%)\end{tabular}    & RMSE     & \begin{tabular}[c]{@{}c@{}}Coverage\\ (\%)\end{tabular}     & RMSE     & \begin{tabular}[c]{@{}c@{}}Coverage\\ (\%)\end{tabular}     & RMSE     \\ \hline
$\alpha_{1}$ & 100         & 0.004    & 100          & 0.003    & 0.96         & 0.004    \\
$\beta_{1,1} $ & 0.94        & 0.02     & 0.9          & 0.02     & 0.96         & 0.02     \\
$\beta_{1,2} $ & 0.98        & 0.03     & 0.92         & 0.03     & 0.92         & 0.03     \\
$\alpha_{2}$ & 0.98        & 0.003    & 0.92         & 0.004    & 0.98         & 0.004    \\
$\alpha_{3}$ & 0.98        & 0.004    & 0.94         & 0.004    & 0.88         & 0.005    \\ \hline
\end{tabular}
\caption{Coverage rates for $95\%$ confidence interval enclosing the true value and RMSE for estimated coefficients, across 50 replications for Simulation Regime 5.}
\label{covsim5}
\end{table}
\newpage
\section{Additional details for \gr{the} real data application}

In addition to the Autoregressive model with lag 8 (AR(8)) fitted individually to each time series in Section 6 of the main article, we present in Table \ref{rmse_ar} results from the predictive performance of the AR model for a range of lags by reporting the RMSE of the prediction.

\begin{table}[htb!]
\centering
\begin{tabular}{cc}
\multicolumn{2}{c}{AR} \\ \hline
lag      & RMSE        \\ \hline
1        & 0.9423      \\
2        & 0.9365      \\
3        & 0.9381      \\
4        & 0.9435      \\
5        & 0.9433      \\
6        & 0.9433      \\
7        & 0.9433      \\ \hline
\end{tabular}
\caption{RMSE from fitting an AR model for various lag sizes.}
\label{rmse_ar}
\end{table}